\documentclass[referee]{raa_rb}
\usepackage{graphicx,times}
\usepackage{natbib}
\usepackage{amssymb,amsmath}
\bibpunct{(}{)}{;}{a}{}{,}

\usepackage[a4paper=true,dvipdfm=true,pagebackref=true]{hyperref}
\hypersetup{pdftitle = The title of my PDF, pdfauthor = My name, pdfsubject= The subject, pdfkeywords = keyword1 keyword2 keyword3}
\hypersetup{colorlinks = true, linkcolor = green, anchorcolor = red, citecolor = blue, filecolor = red, pagecolor = red, urlcolor = red}

\begin{document}

   \title{Line Profile studies of Hydrodynamical Models of Cometary Compact H II Regions
\footnotetext{ }
}

 \volnopage{ {\bf 2012} Vol.\ {\bf X} No. {\bf XX}, 000--000}
   \setcounter{page}{1}

   \author{Feng-Yao Zhu\inst{1,2}, Qing-Feng Zhu\inst{1,2}
   }

   \institute{ Department of Astronomy, University of Science and Technology of China, Hefei 230026,
China; {\it zhufya@mail.ustc.edu.cn} {\it zhuqf@ustc.edu.cn}\\
        \and
Key Laboratory for Research in Galaxies and Cosmology, University of Science and Technology of China, Chinese Academy of Sciences, Hefei, Anhui, 230026, China\\
\vs \no
   {}
}

\abstract{We simulate evolution of cometary H II regions based on several champagne flow models and bow shock models, and calculate the profiles of the [Ne II] fine-structure line at $12.81\mu m$, the $H30\alpha$ recombination line and the [Ne III] fine-structure line at $15.55\mu m$ for these models at different inclinations of $0^o,~30^o~\textrm{and}~60^o$. We find that the profiles in the bow shock models are generally different from those in the champagne flow models, but the profiles in the bow shock with lower stellar velocity ($\leq5km~s^{-1}$) are similar to those in the champagne flow models. In champagne flow models, both the velocity of peak flux and the flux weighted central velocities of all three lines are pointing outward from molecular clouds. In bow shock models, the directions of these velocities rely on the speed of stars. They have the similar motion in high stellar speed case but opposite directions in low stellar speed case. We notice that the line profiles from the slit along the symmetrical axis of the projected 2D image of these models are useful for distinguishing bow shock models and champagne flow models. It is also confirmed by the calculation that the flux weighted central velocity and the line luminosity of the [Ne III] line can be estimated from the [Ne II] line and the $H30\alpha$ line.
}

   \authorrunning{F.-Y. Zhu \& Q.-F. Zhu}            
   \titlerunning{Line profiles from the H II region}  
   \maketitle

%
\section{Introduction}           
\label{sect:intro}

Massive stars always form in dense molecular clouds and transfer a great amount of energy from their ionizing fluxes and stellar winds to the surrounding interstellar medium (ISM). This affects the kinetic and thermal energy of the molecular clouds so as to change the velocity structure and the morphologies of the clouds substantially. The ionizing photons from a massive star can ionize the surrounding ISM and form an H II region. In a uniform environment, the H II region will be spherical and is called Str\"{o}mgren sphere \citep{str39}. However, observations show that a large number of young H II regions have a cometary morphology \citep{woo89,kur94,wal98}. \citet{ten79} pointed out that these cometary H II regions result from the density gradients in the molecular cloud. In this explanation, the comet-shaped H II regions are called blister H II regions or named as champagne flows. The Orion Nebula is the archetype of this kind of H II regions. Observation shows that there is a bright ionization front on the surface of a molecular cloud and ionized gas flows away from the cloud \citep{isr78}. The champagne flow model was first created by Tenorio-Tagle and coworkers \citep{ten79,bod79,yor83}. In these studies, the H II region is assumed to form in a uniform molecular cloud, but close to the boundary of it. The H II region will expand and break out into the intercloud medium of low density to form a shock. The Mach numbers of the resulting shocks are shown to be $30-42$, and the velocities of the ionized gas can reach up to $40~km~s^{-1}$. Since then, a series models were developed by including the effect of a stellar wind, a magnetic field or an exponential density distribution \citep{com97,art06,gen12}. In all of these models, a density gradient is assumed to exist, which is considered as the main cause of the cometary morphology.

There is an alternative explanation to the cometary shape of H II regions called bow shock. In this kind of models, a wind-blowing ionizing star moves supersonically with respect to the dense molecular cloud, and forms a shock in front of the star. A cometary H II region can also form in this case \citep{mac91,bur92}. \citet{wil96} created an analytic model to derive the bow shock characteristics. Other researchers made a series numerical models to simulate bow shocks \citep{com98,art06}. The presence of the density gradient seems unavoidable in the non-homogeneous environment in molecular clouds. High speed stars are not common in the Galaxy. But, it can happen in the region with a high stellar density, such as massive star forming regions. Hence, it is necessary to compare the two possible causes of cometary H II region to interpret observations.

\citet{art06} pointed out that simple champagne flows without a stellar wind will not show a limb-brightened morphology. By contrast, bow shock models always display this morphology. They also pointed out that the two kinds of models can be distinguished in the kinematics. For example, the highest velocity of ionized gas with respect to the molecular clouds is at the head of the cometary H II region in the bow shock models while it exists at the tail in a champagne flow model. It is the most notable that the directions of motion of the ionized gas at the head are always toward the molecular clouds in bow shock models but could be away from the clouds in champagne flow models \citep{art06}. Since the profiles of emission lines are partly determined by kinematics, this difference in kinematics could lead to different line profiles. Line profiles are direct observing properties that can be accessed with high resolution observations. Therefore, line profiles can be used as the criterion to select better models for certain observations.

In this paper, we present the profiles of the [Ne II] $12.81 \mu m$ line, the [Ne III] $15.55 \mu m$ line and the hydrogen $H30\alpha$ ($31-30$) recombination line from cometary H II regions by simulating bow shock models and champagne flow models with different parameters. The [Ne II] $12.81 \mu m$ line is the brightest line from HII regions in the mid-infrared, which can be accessed by TEXES on IRTF or EXES on Stratospheric Observatory for Infrared Astronomy (SOFIA). The [Ne III] $15.55\mu m$ line can not be observed from the ground. In order to investigate the profile of the [Ne III] $15.55\mu m$ line, we have to rely on future space telescopes. The $H30\alpha$ line can be observed by Caltech Submillimeter Observatory (CSO). These lines are useful to study the gas kinematics of compact HII regions. We compare the line profiles between the bow shock models and champagne flow models and discuss the possibility to distinguish these two kinds of models based on these line profiles. The organization of this paper is as follows: In \S 2, we describe the method of our numerical simulation. In \S 3, we present the results of the numerical models. \S 4 show our conclusions.


\section{Method}

\subsection{Method of Hydrodynamics, Radiative Transfer and Thermal Processes}

It is necessary to simulate the time evolution of the cometary H II regions in order to obtain the profiles of lines of interest. To correctly describe the time evolution of the regions, one has to treat the transportation of energy properly. In the situation of emission nebula, the gravity is not important, but the hydrodynamics and the radiative transfer are essential processes for energy transportation. In this paper, both hydrodynamics and radiative transfer are considered.

A 2D explicit Eulerian hydrodynamic method is used to treat the evolution on a cylindrically symmetric grid. Most of the models in this work are computed on a $250\times500$ grid. A big grid of $400\times400$ is used when computing blister H II region model to check the contribution of the ionized gas far from the star at the sides to the emission line profile. The results show that the grid of $250\times500$ is enough. The cell size of the grids is chosen to be $dr=0.005~pc$. A HLLC Riemann solver \citep{miy05} is used to solve the hydrodynamic conservation equations.

When treating the radiative transfer, we consider the star as a single source for ionizing and dissociating radiation. We solve the radiative transfer for EUV ($hv\geq13.6eV$) and FUV ($11.26eV\leq hv<13.6eV$), respectively \citep{dia98}. The black body spectrum is assumed for the ionizing star. We assumed the "on the spot" approximation when treating ionizing photons. For dissociating radiation, because the column density of molecular hydrogen exceeds $10^{14} cm^{-2}$ in our models, the FUV lines are optically thick. Thus, the self-shielding by $H_2$ becomes important \citep{hol99}. The dissociation rate and reformation rate of hydrogen molecules are calculated by using a simple self-shielding approximation introduced in \citet{dra96} and the method in \citet{hol99}, respectively. The dissociation and reformation of CO molecules are also included in the models, following the methods given by \citet{lee96} and \citet{nel97}.

In the ionized region, the photoionization heating is considered as the only heating process \citep{spi78}. We use the cooling curve for solar abundances given by \citet{mel02} to compute the radiative cooling rate. This cooling curve is derived on the assumption that the cooling of the gas is due to collisional excitation of hydrogen and metal lines and hydrogen recombination. Outside of the H II region, the heating processes of the gas include photoelectric heating, heating from photodissociation, reformation of hydrogen, cosmic ray and FUV pumping of $H_2$ molecules as heating processes. The following cooling processes are considered: atomic fine-structure lines of [O I] $63\mu m$, [O I] $146\mu m$ and [C II] $158\mu m$, the rotational and vibrational transitions of $CO$ and $H_2$, dust recombination and gas-grain collisions \citep{hol79,tie85,hol89,bak94,hos06}. Since the results of the photodissociation region will not be presented in this paper, the purpose of including the radiative transfer of the photodissociation radiation is mainly to conserve the energy and momentum of the gas.

\subsection{Line profiles}

After we solve the continuum, momentum and energy equations of the hydrodynamic models, the method derived by \citet{gla07} is applied to compute the line luminosity for the [Ne II], [Ne III] and $H30\alpha$ lines. The line luminosity $L$ at a given velocity $v$ is as follow:

\begin{equation}
L(v)=\frac{1}{\sqrt{2\pi}v_{th}}\int \textrm{exp}(-\frac{[v-v_{los}(\textbf{r})]^2}{2v_{th}^2})f_{ul}(i)dV~,      \\
\end{equation}

\begin{equation}
f_{ul}(i)=\begin{cases}
b_kN_kA_{ul}h\nu_{ul} & for~H30\alpha,~k=31,~i=0  \\
Ab_{Ne}X(Ne^+)nP_uA_{ul}h\nu_{ul} & for~\textrm{[Ne II]},~i=1 \\
Ab_{Ne}X(Ne^{2+})nP_uA_{ul}h\nu_{ul} & for~\textrm{[Ne III]},~i=2

\end{cases}~~~~,
\end{equation}
where $v_{los}(\textbf{r})$ is the line-of-sight component of the velocity vector. $h\nu_{ul}$ is the photon energy. $v_{th}$ is the thermal velocity of gas. $b_kN_k$ is the number density of hydrogen atoms in the kth electronic energy level ($k=31$ for $H30\alpha$). $N_k$ is the theoretical value for the number density in the kth level expected in LTE and is proportional to $n_e^2$. $b_k$ is departure coefficient \citep{sea59}. $n$ is the number density of $H$ nuclei. The abundance of $Ne$, $Ab_{Ne} = 1.0\times10^{-4}$, is adopted in these models \citep{hol01}. $A_{ul}$ is the Einstein emission coefficient for corresponding transitions \citep{ale08,gla07}. $X(Ne^+)$ and $X(Ne^{2+})$ are the fraction of $Ne^+$ and $Ne^{2+}$ ions, respectively. $P_u$ is the excitation fraction of the upper state and is computed as in \citet{gla07}:

\begin{equation}
P_u=\begin{cases}
[2(1+n_{cr}/n_e)exp(1122.8/T)+1]^{-1} & for~\textrm{[Ne II]}  \\
[1+(5/3)(1+n_{cr1}/n_e)exp(925.3/T)+(1/3)(1+n_{cr2}/n_e)exp(-399/T)]^{-1} & for~\textrm{[Ne III]}

\end{cases}~~~~.
\end{equation}
where the critical density $n_{cr}=5.53\times10^3T^{0.5}~cm^{-3}$, $n_{cr1}=3.94\times10^3T^{0.5}~cm^{-3}$ and $n_{cr2}=7.2\times10^2T^{0.5}~cm^{-3}$. The relative fractions of $Ne$, $Ne^+$ and $Ne^{2+}$ ($X(Ne)+X(Ne^+)+X(Ne^{2+})=1$) are computed through the ionization-recombination balance equations. The photoionzation cross sections, the recombination rate coefficients and the charge exchange rate coefficient are given by \citet{hen70}, \citet{peq91} and \citet{gla07}, respectively.
The collisional ionization of $Ne$ is not considered in photoionized region where temperature is much lower than the critical temperature $T_{c}=2.50\times10^5~K$, and the gas density is relatively low. The contribution of the hot stellar wind bubble ($T=10^{6-8}~K$) to line luminosities is negligible because of its low density ($n<5~cm^{-3}$). Hence, collisional ionization of neon atoms is not considered. In addition, since the average energy of photons from young stars are not high energy, the fractions of highly ionized ions as $Ne^{3+}$ and $Ne^{4+}$ are generally not considered \citep{mor02}. This conclusion is also suggested by our calculations. We have tested calculating the relative fractions of $Ne^{3+}$ ions and other higher ionization state $Ne$ ions. They are always lower than $0.002$ in photoionized region, and can be safely neglected. X rays and EUV photons with energies greater than 21.56eV can both photoionize neon, but we neglect X-ray ionization because of the lack of x ray sources relative to EUV photons. The results of the $H30\alpha$ recombination line can be applied to other $H$ recombination lines. They have the same normalized profile when the pressure broadening effect can be neglected.

In this paper, we assume that the observers are viewing the cometary regions from the tail to the head. Therefore, a blue-shifted velocity suggests that the gas mainly moves at a direction from the head to the tail and a red-shifted velocity suggests the gas mainly moves at the opposite direction.

\section{Results}

In this section, seven models are presented. Four of them are bow shock models, and the rest are champagne flow models. The parameters of these models are presented in Table \ref{tab_modc}. These models are selected to test the effects of different model ingredients on the line profiles. The values of stellar parameters are based on \citet{dia98} and \citet{dal13}. The initial density is consistent with the condition in compact H II region. In the following section, the models will be described and analyzed individually.

\begin{table}[!htp]\footnotesize
\centering
\begin{tabular}{l|llll|ll}
\hline
Model & $\dot{M}(M_\odot yr^{-1}) $ & $v_\ast(km~s^{-1})$ & $v_w (km~s^{-1})$ & $log[S_{UV}(s^{-1})]$ & $n_0(cm^{-3})$ & Scale Height(pc)   \\
\hline
A & $9.93\times10^{-7}$ & 10 & $2720.1$ & 48.78 & 8000 & 0 \\
B & $9.93\times10^{-7}$ & 0 & $2720.1$ & 48.78 & 8000 & 0.05 \\
\hline
C & $9.93\times10^{-7}$ & 0 & $2720.1$ & 48.78 & 8000 & 0.15 \\
\hline
D & $9.93\times10^{-7}$ & 15 & $2720.1$ & 48.78 & 8000 & 0 \\
E & $9.93\times10^{-7}$ & 5 & $2720.1$ & 48.78 & 8000 & 0 \\
\hline
F & $3.56\times10^{-7}$ & 0 & $1986.3$ & 48.10 & 8000 & 0.05 \\
G & $3.56\times10^{-7}$ & 10 & $1986.3$ & 48.10 & 8000 & 0 \\
\hline
\end{tabular}
\caption{The model parameters in model A-G.\label{tab_modc}}
\end{table}

\subsection{Model A}

In Model A, we simulate a stellar bow shock in a uniform medium with the number density of $n_0=8000~cm^{-3}$. The velocity of the moving star is assumed to be $v_\ast=10~km~s^{-1}$. When simulating bow shock models, we carry out the calculation in the rest frame of the star first, so that the same procedures can also be used in the champagne flow cases. After the calculation, we convert all velocities to the values in the rest frame of molecular clouds. The velocities presented in this paper are all in the frame of reference of molecular clouds. In all models, the z-axis is parallel to the symmetrical axis and the positive direction is from the tail to the head of the cometary region. The star is at the position of $(x,z)=(0,0)$. The effective temperature of the star is $40,000K$. The numbers of ionization photons ($h\upsilon \geq 13.6eV$) and the photodissociation photons ($11.26\leq h\upsilon<13.6eV$) emitted from the star per second are $10^{48.78}s^{-1}$ and $10^{48.76}s^{-1}$, respectively. The mass-loss rate is $\dot{M}=9.93\times10^{-7} M_\odot yr^{-1}$, and the terminal velocity of the stellar wind is $v_w=2720.1~km~s^{-1}$. These parameters are consistent with a star of mass $M_\ast=40.9~M_\odot$ \citep{dia98,dal13}.

\begin{figure}[!htp]
\centering
\includegraphics[scale=.45]{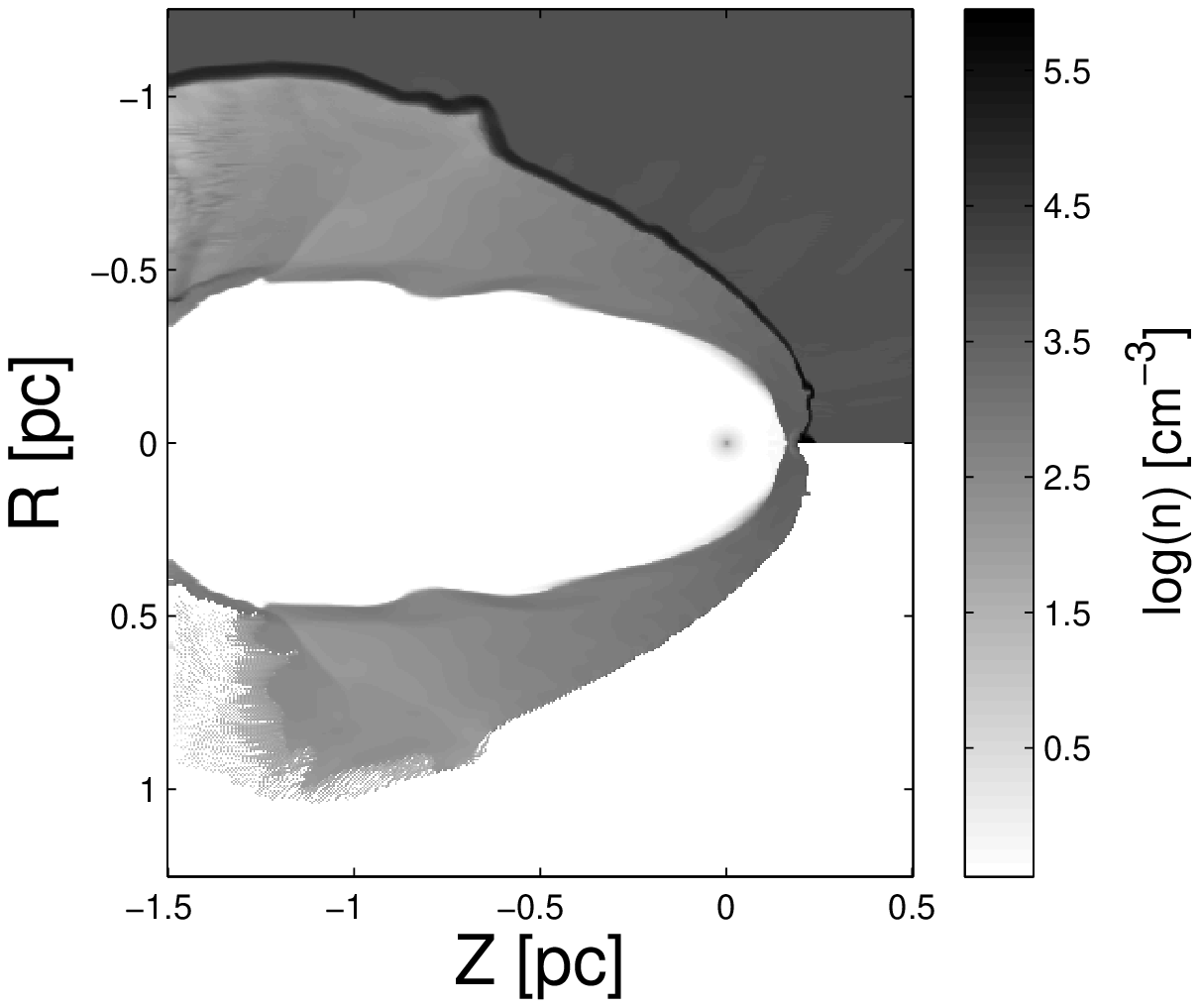}
\includegraphics[scale=.45]{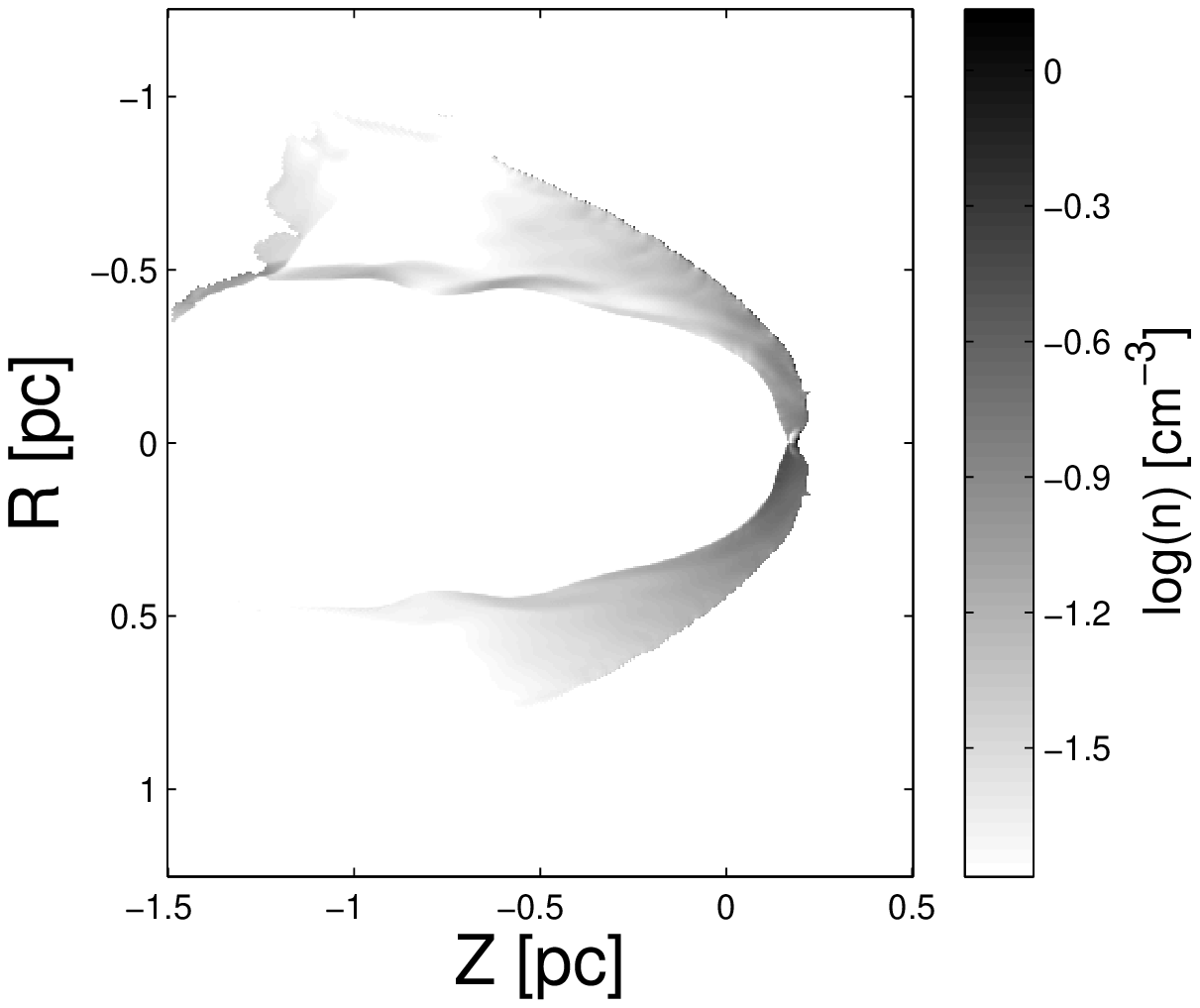}
\caption{Gas density at logarithmic scales at the age of $120,000~yrs$. In the left panel, the top half of the figure shows the total density of the gas and the bottom half shows the density of $H^+$ ions. The right panel shows the densities of $Ne^+$ ions (top half) and $Ne^{2+}$ ions (bottom half).}
\begin{flushleft}
\end{flushleft}
\label{fig_moda1}
\end{figure}

Our simulation is stopped at $120,000~yr$. Before that time, the ionization front ahead of the star has been approximately motionless relative to the star for a few $10^4$ years. In Figure \ref{fig_moda1}, the number density of all materials, $H^+$, $Ne^+$ and $Ne^{2+}$ ions in model A are presented. A stellar wind bubble of low density ($n<5~cm^{-3}$) can be seen around the star. The bubble is surrounded by the photoionizated region of high density ($n\sim200-20000~cm^{-3}$), and a dense neutral shell ($n\sim10^{5-6}~cm^{-3}$). In the right panel of Figure \ref{fig_moda1}, the density distributions of $Ne^+$ and $Ne^{2+}$ ions are shown in grey scales. the $Ne^+$ and $Ne^{2+}$ ions distribute in the whole H II region, but we only show the number density higher than $10^{-1.8}~cm^{-3}$ in order to highlight the different distributions between $Ne^+$ and $Ne^{2+}$ ions. Although the high densities in the head of the H II region, which is defined as the ionized region of $z>0$, leads to the high densities of both $Ne^+$ and $Ne^{2+}$ there, the density of $Ne^+$ ions roughly increases with the distance from the star in the head of the H II region. The density of $Ne^{2+}$, by contrast, decreases with the distance in the head region. Large fluxes of the photons with energy higher than the ionization potential of $Ne^+$ can increase the ionization rate from $Ne^+$ into $Ne^{2+}$ at the locations near the star. On the contrary, high electron density in the head region would increase the recombination rate. This leads to the different density distributions of $Ne^+$ and $Ne^{2+}$ ions. 

\begin{figure}[!htp]
\centering
\includegraphics[scale=.45]{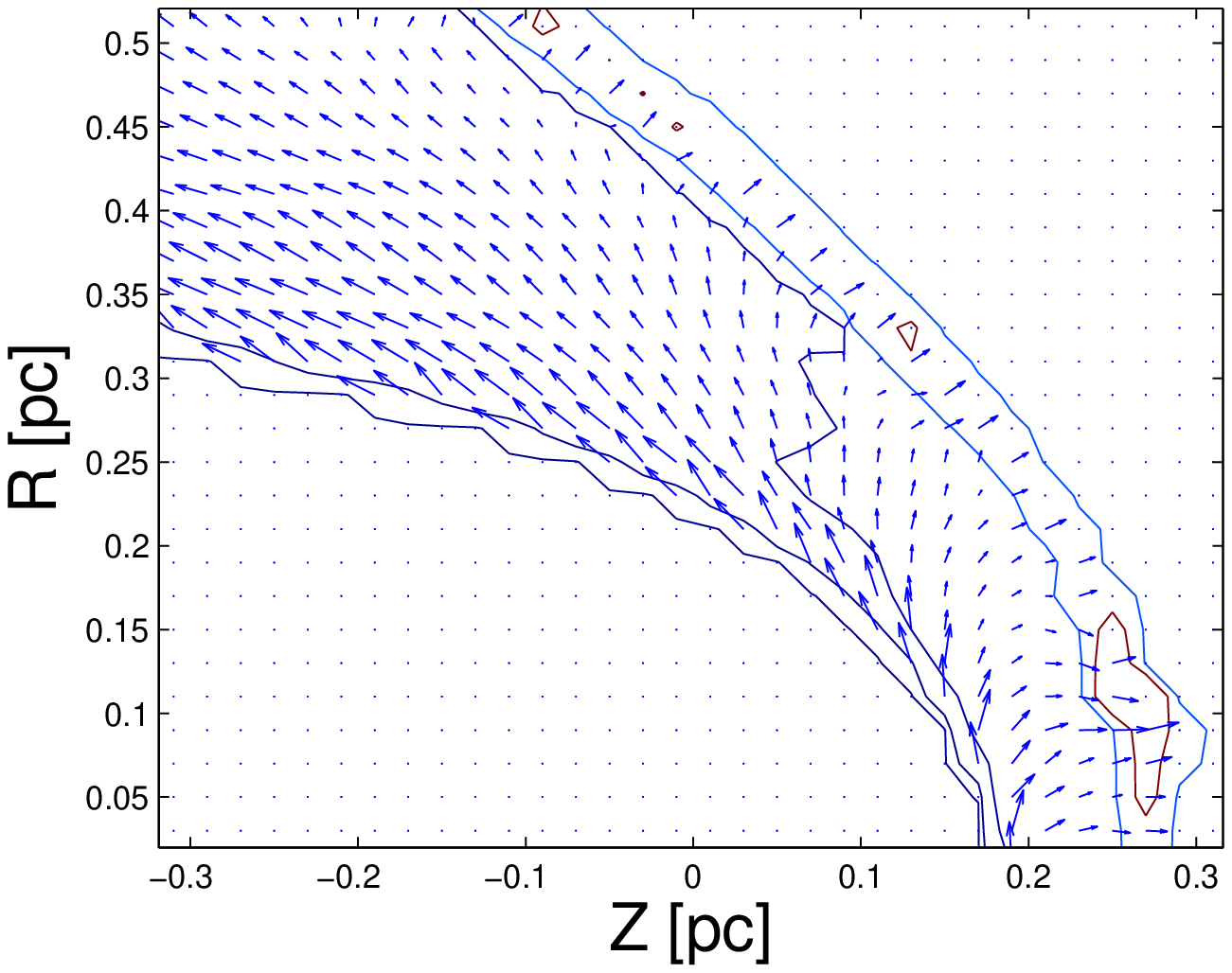}
\includegraphics[scale=.45]{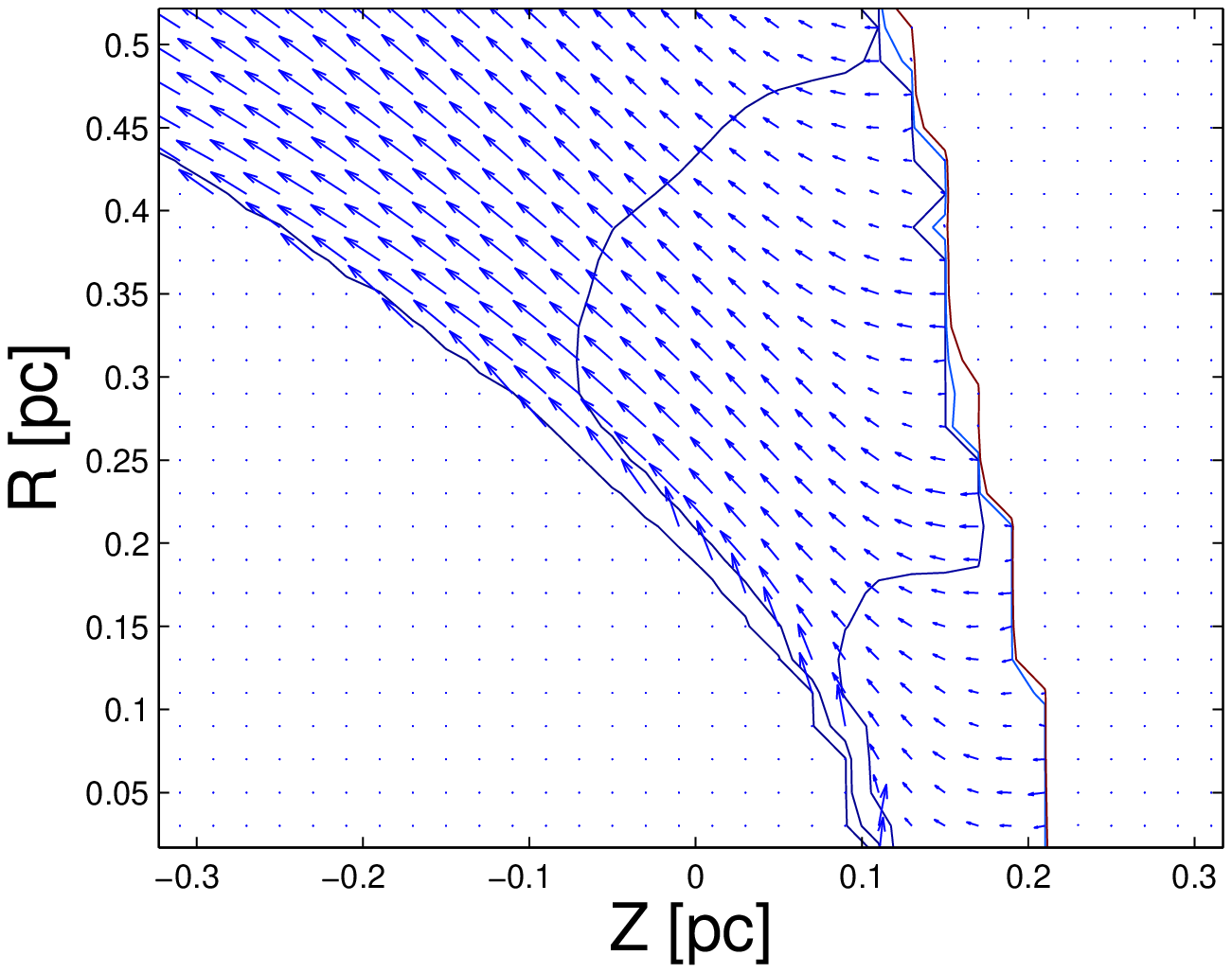}
\caption{The velocity fields of photoionized gas in model A (bow shock, left) and model B (champagne flow, right). The velocity fields in the stellar wind bubble($n < 50~cm^{-3}$) and outside the H II region ($X(H^+) < 0.1$) are not shown. The arrows represent velocities of $v > 1~km~s^{-1}$, and the lengths of them are proportional to their absolute values. The position of the star is (0,0). The five contour levels are at $50$, $500$, $2000$, $20000$ and $100000~cm^{-3}$.}
\begin{flushleft}
\end{flushleft}
\label{fig_discus}
\end{figure}

\begin{figure}[!htp]
\centering
\includegraphics[scale=.33]{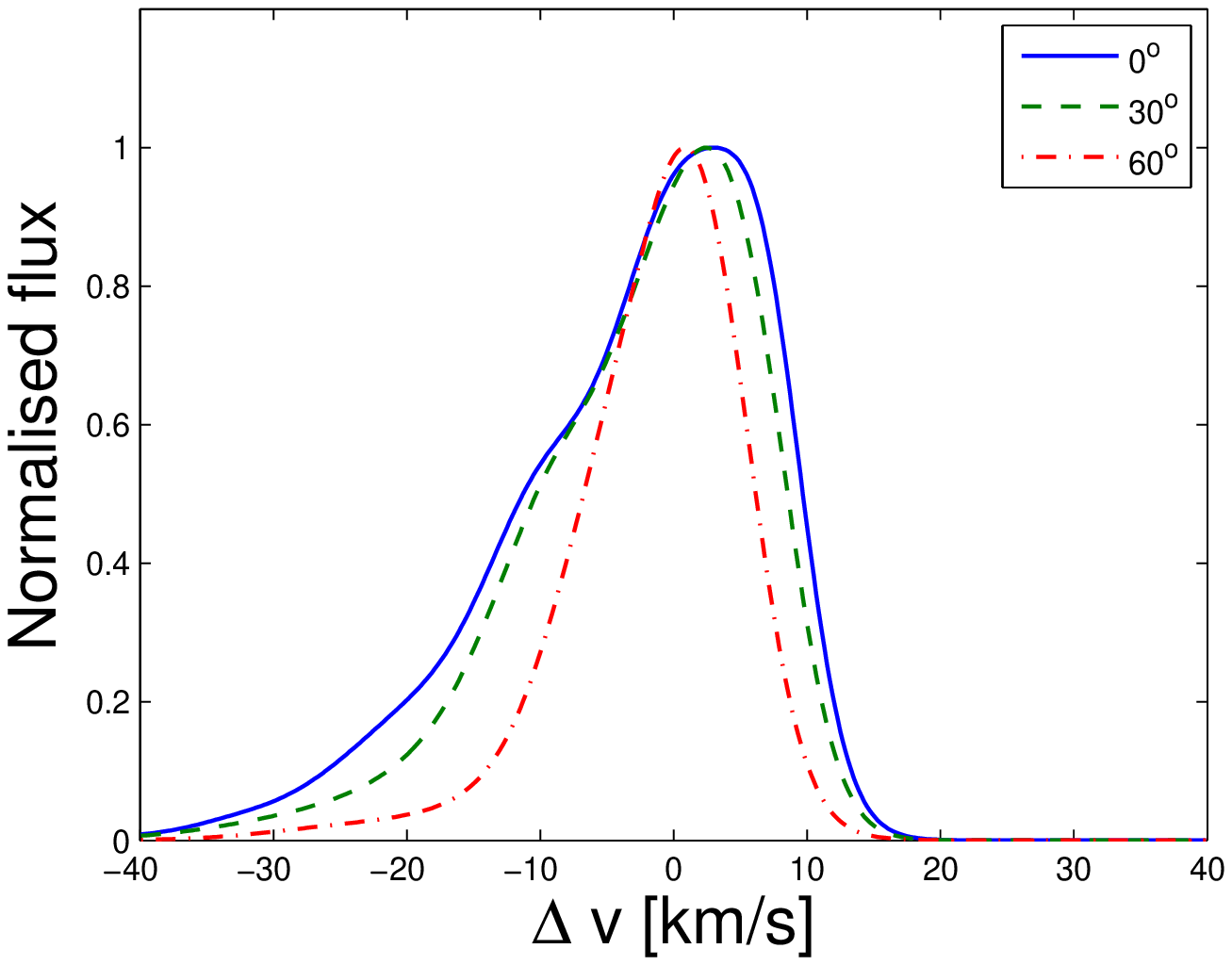}
\includegraphics[scale=.33]{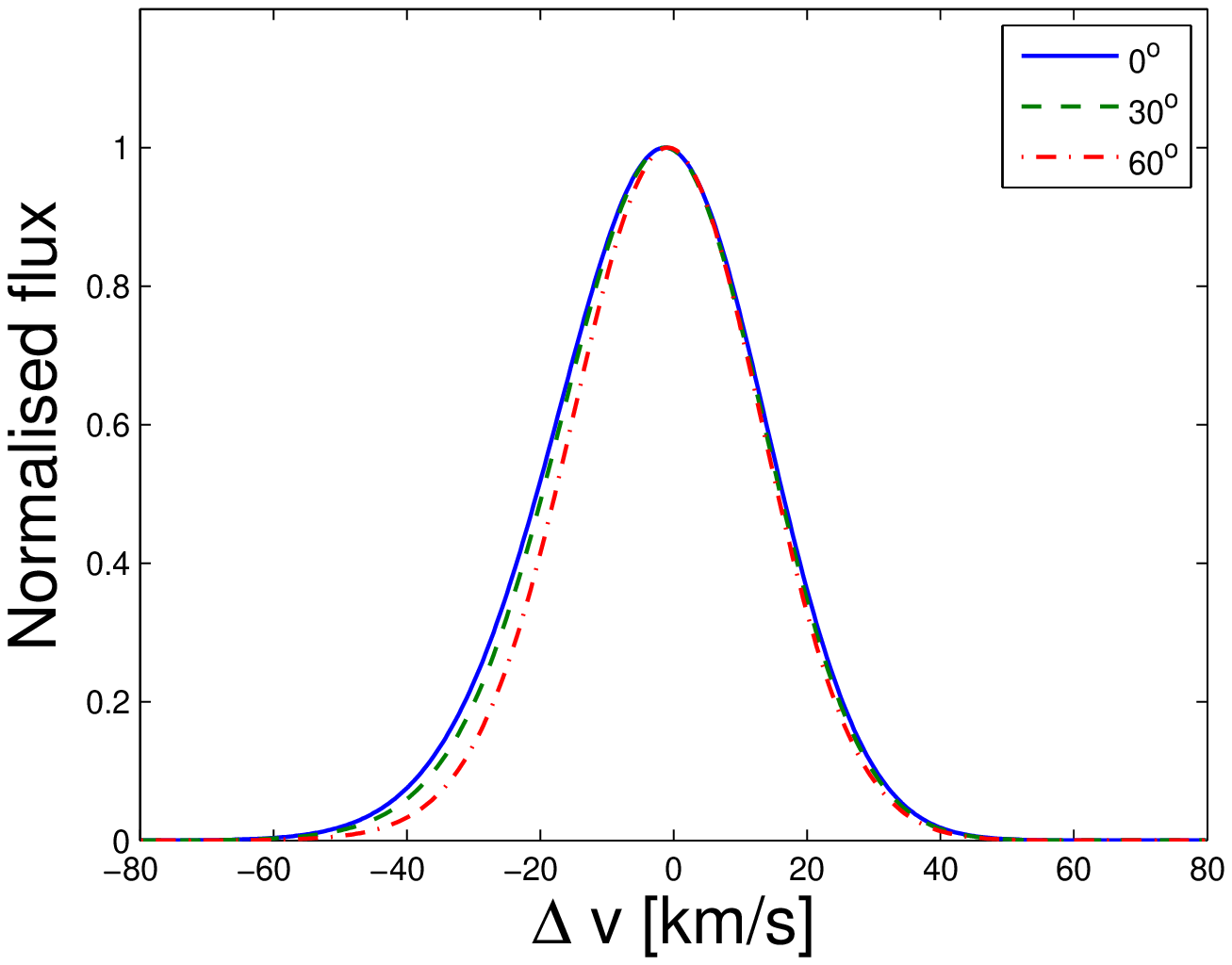}
\includegraphics[scale=.33]{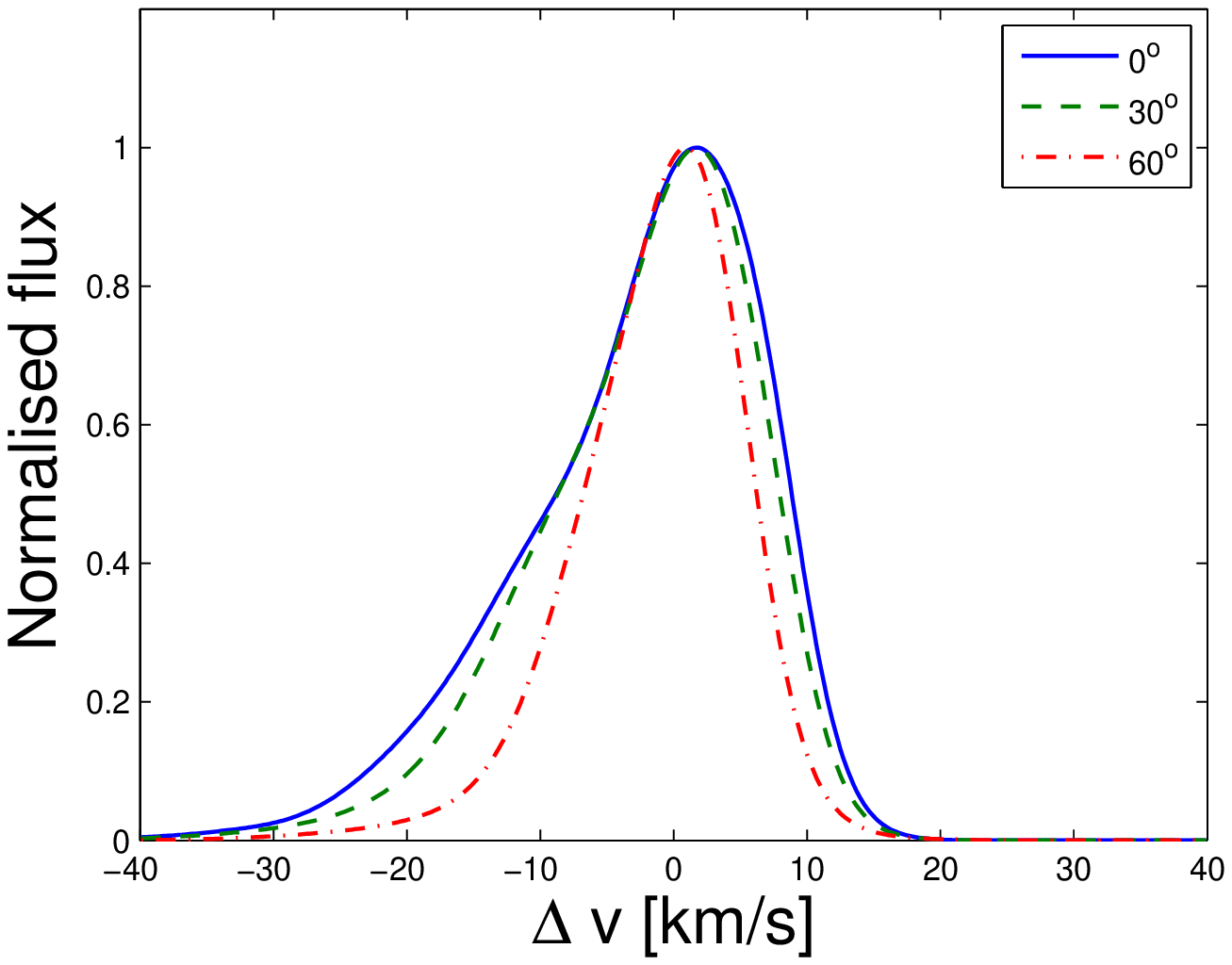}
\caption{profiles of the [Ne II] $12.81 \mu m$ line (left), the $H30\alpha$ line (middle) and the [Ne III] $15.55 \mu m$ line (right) from the cometary H II regions in model A for three inclination angles ($0^o,~30^o,~60^o$).}
\begin{flushleft}
\end{flushleft}
\label{fig_moda}
\end{figure}

\begin{table}[!htp]\footnotesize
\centering
\begin{tabular}{l|llll|l}
\hline
Line & Inclination & Peak & FWHM & FWCV  & Luminosity \\
 & & ($km~s^{-1}$) & ($km~s^{-1}$) & ($km~s^{-1}$) & ($erg~s^{-1}$) \\
\hline
[Ne II] $12.81 \mu m$ & $0^o$ & 3.1 & 20.8 & -3.32 & $7.75\times10^{35}$ \\
 & $30^o$ & 2.5 & 19.0 & -2.92 \\
 & $60^o$ & 0.9 & 13.0 & -1.70 \\
\hline
$H30\alpha$ & $0^o$ & -1.4 & 36.8 & -3.40 & $1.07\times10^{30}$ \\
 & $30^o$ & -1.4 & 35.6 & -2.99 \\
 & $60^o$ & -1 & 33.2 & -1.74 \\
\hline
[Ne III] $15.55 \mu m$ & $0^o$ & 1.7 & 17.6 & -3.28 & $1.03\times10^{36}$ \\
 & $30^o$ & 1.7 & 16.6 & -2.85 \\
 & $60^o$ & 0.9 & 13.0 & -1.64 \\
\hline
\end{tabular}
\caption{Peaks, FWHMs and flux weighted central velocities(FWCV) of lines at different angles in model A\label{tab_moda}}
\end{table}

\begin{figure}[!htp]
\centering
\includegraphics[scale=.35]{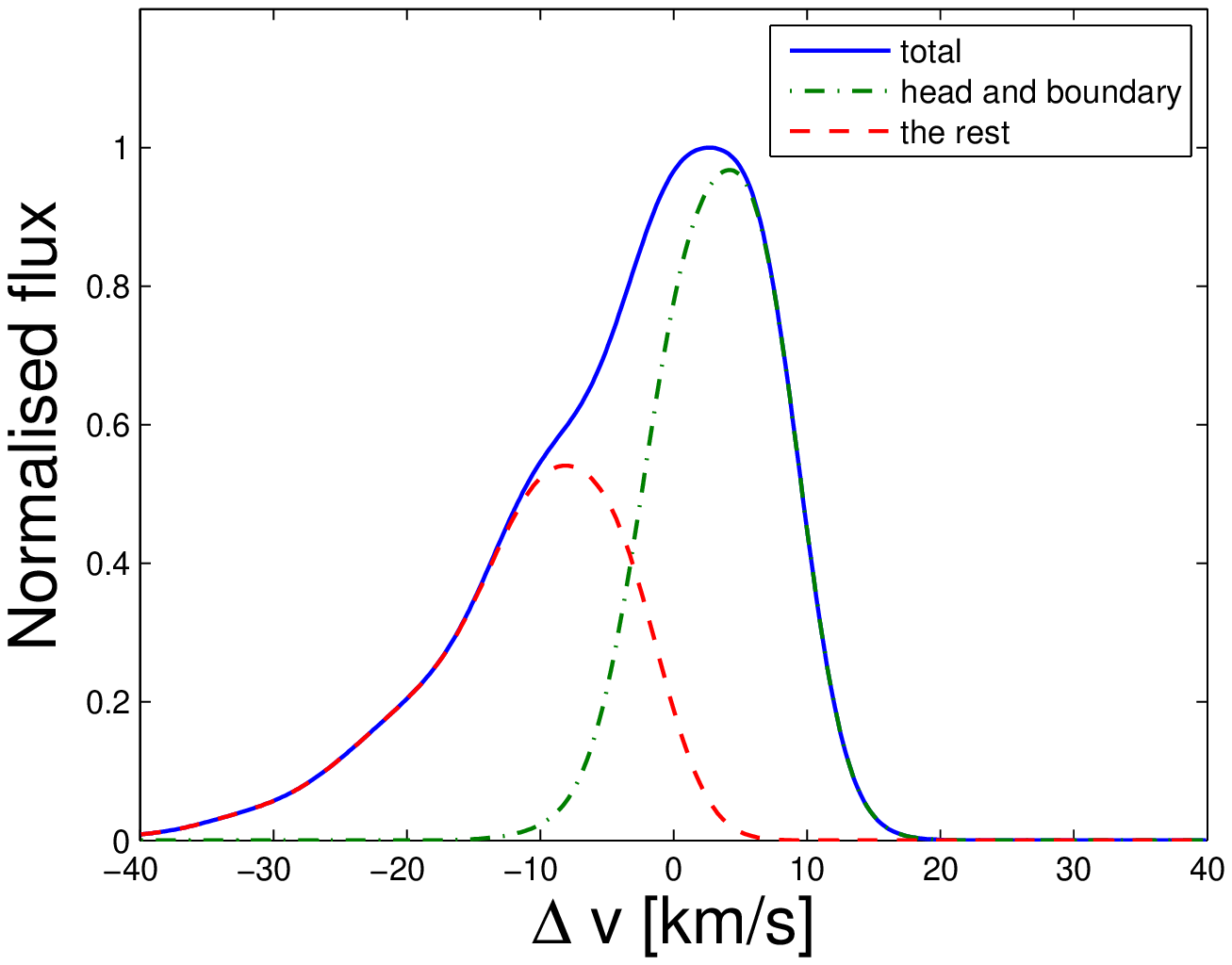}
\includegraphics[scale=.35]{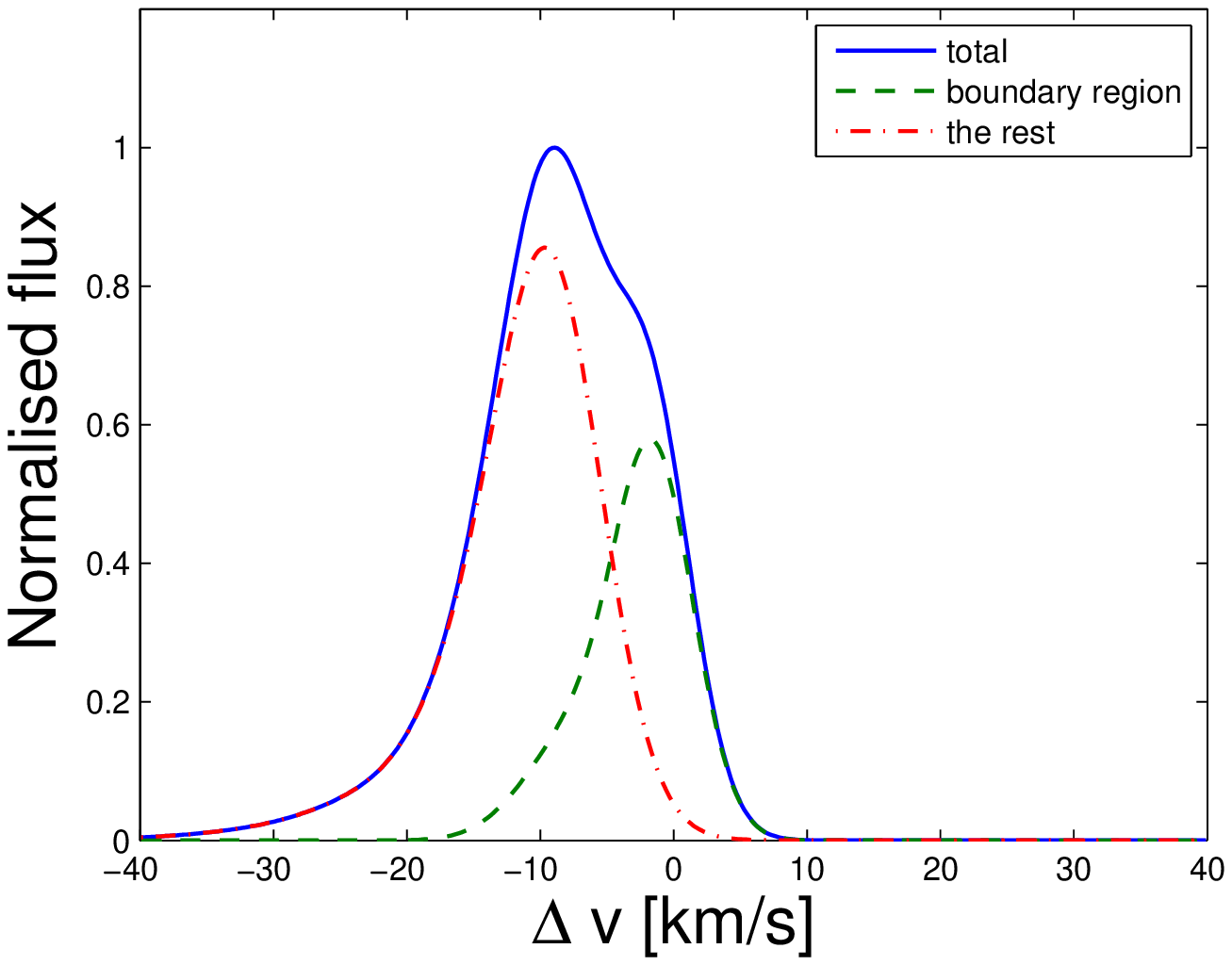}
\caption{the line profiles of the [Ne II] $12.81 \mu m$ from the whole H II region, the head and boundary region and the rest part of the H II region. The left panel is for model A, and the right panel is for model B.}
\begin{flushleft}
\end{flushleft}
\label{fig_modals}
\end{figure}

In Figure \ref{fig_discus}, the velocity field of photoionized gas is shown. The result of the bow shock (model A) is shown in the left panel. A large proportion of ionized gas in the head of the H II region has red-shifted velocities at the direction from the tail to the head, and the velocities of the ionized gas in the boundary region, which is defined as a $0.025pc$ wide layer in the H II region near the ionization front, are approximately perpendicular to the ionization front. Particularly, the gas in front of the star moves at the similar direction as the stellar motion. The directions of motions of the ionized gas in the rest part of the H II region are gradually turned toward the tail. This phenomenon is because of the accelerations toward the tail due to the pressure density in the H II region. The details of this reason is discussed in \S \ref{sect:stellar motion}.

In Figure \ref{fig_moda}, the profiles of the [Ne II] line, the $H30\alpha$ line and the [Ne III] line from the H II region for different inclinations ($\theta=0^o,~30^o,~\textrm{and}~60^o$ from the z-axis ) are presented. The profiles are all asymmetrical, but the $H30\alpha$ line is more symmetric than other two lines. Both the [Ne II] line and the [Ne III] line profiles are skewed to the right and have a long tail at the left side. With increasing inclination angles, these line profiles become less asymmetrical and narrower. It is obvious that the $H30\alpha$ line is much broader than the [Ne II] line and the [Ne III] line. This is due to the larger thermal broadening of lower mass hydrogen.

In Table \ref{tab_moda}, the peak locations, FWHMs and flux weighted central velocities (FWCV) of the lines for three inclinations of $0^o$, $30^o$ and $60^o$ are shown. For the [Ne II] line, the peak locations and FWHMs of the profiles decrease with increasing inclination angles. Although the peak locations are all red-shifted ($3.1,~2.5~and~0.9~km~s^{-1}$), the flux weighted central velocities are all blue-shifted ($-3.32,~2.92~and~-1.70~km~s^{-1}$). This suggests that the contributions of blue-shifted tail in the profile are significant. Because the line profile for the inclination of $0^o$ seem to be consist of two components, we calculate the [Ne II] line profile from the head and the boundary region and the line profile from the rest part of the H II region, separately. These two line profiles are plotted in Figure \ref{fig_modals}. It can be seen that the emission from the head region and the boundary region leads to the red-shifted peak location of the total line profile, and and the gas from the rest part of the H II region mainly contributes to the blue-shifted tail of the line profile. The luminosity of the [Ne II] line is $7.75\times10^{35}~erg~s^{-1}$ which is independent of inclinations.

For the [Ne III] line, the peak locations are slightly red-shifted ($1.7,~1.7~and~0.9~km~s^{-1}$), and the flux weighted central velocities are all blue-shifted ($-3.4,~-2.99~and~-1.74~km~s^{-1}$). The reason of this is the same as in the [Ne II] line profile. The line luminosity of the [Ne III] line is $1.03\times10^{36}~erg~s^{-1}$. This value is larger than that of the [Ne II] line. For the $H30\alpha$ line, both the peak locations ($-1.4,~-1.4~and~-1.0~km~s^{-1}$) and the flux weighted central velocities ($-3.28,~-2.85~and~-1.64~km~s^{-1}$) are blue-shifted. This is due to the large thermal broadening of lower mass hydrogen. By increasing the broadening, the peak locations of all lines will approach the flux weighted central velocities. The blue-shift of the flux weighted central velocities of the hydrogen line suggests the blue-shifted gas motion dominates for the ionized gas in bow shock model. It seems odd that the ionized gas moves mainly at a direction opposite to the direction of the star and the shock structure. The reason is also the accelerations toward the tail due to the pressure density in the H II region. The luminosity of the $H30\alpha$ line is $1.07\times10^{30}~erg~s^{-1}$ in model A. This is much lower than the luminosities of the [Ne II] line and the [Ne III] line.

\subsection{Model B}
\label{sect:model B}

In model B, the evolution of a champagne flow including a stellar wind is simulated. The density distribution follows an exponential law as $n(z)=n_0exp(z/H)$. $z$ is the axial coordinate along the symmetrical axis and towards the center of molecular cloud. The density at the position of the motionless massive star ($z=0$) is $n_0=8000~cm^{-3}$ initially and the scale height is $H=0.05~pc$. The parameters of the massive star and the stellar wind are the same as in model A. In model B, the simulation of the time evolution is ceased at $160,000~yr$ when the line profiles are roughly stable. The H II region and the neutral region should reach approximate pressure equilibrium, and the champagne flow has completely cleared the low-density material from the grid at the time.

\begin{figure}[!htp]
\centering
\includegraphics[scale=.45]{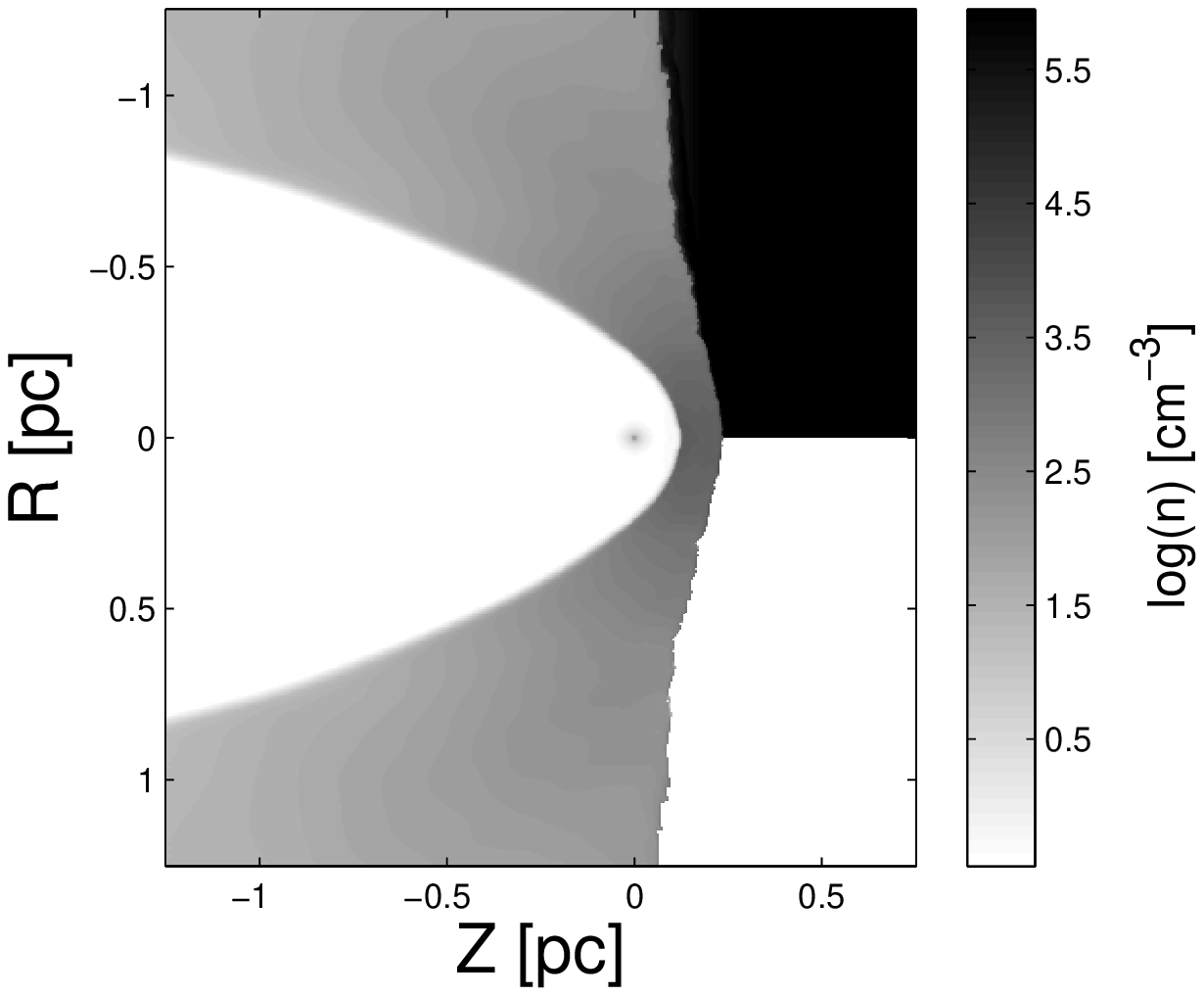}
\includegraphics[scale=.45]{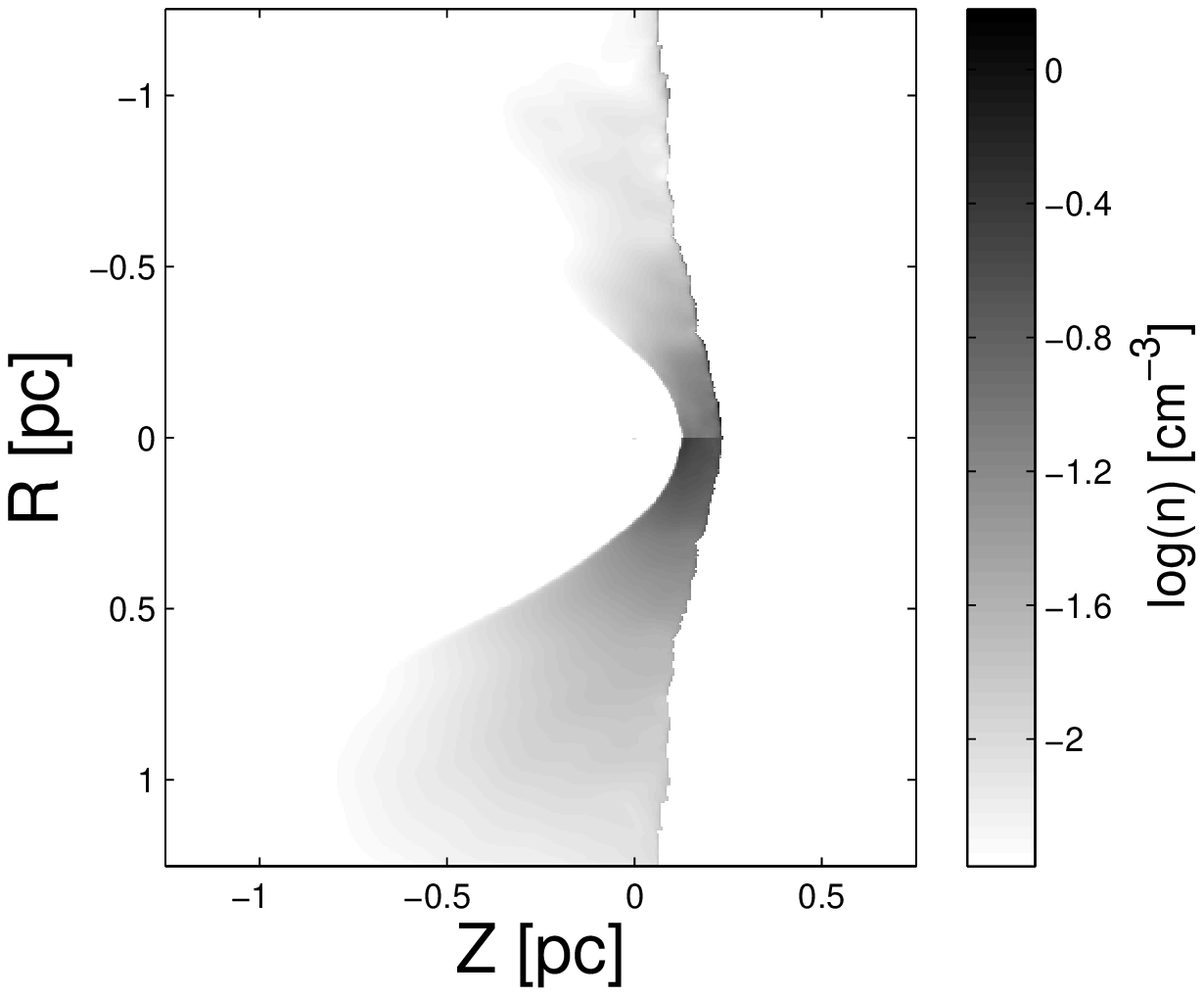}
\caption{Gas density at logarithmic scales at the age of $160,000~yr$ in model B. The top half in the left panel shows the density of all materials. The bottom half in the left panel shows the density of $H^+$ ions. The top half in the right panel shows the density of $Ne^+$ ions. The bottom half in the right panel is for $Ne^{2+}$ ions. The density is in the units of $cm^{-3}$.}
\begin{flushleft}
\end{flushleft}
\label{fig_modb}
\end{figure}

The density distribution in model B at the end of simulation is presented in Figure \ref{fig_modb}. As in model A, there are also a stellar bubble surrounded by the photoionized region and a dense neutral shell which separates the H II region from the dense cloud. The size of H II region in model B is bigger than in model A, but the total number of $H^+$ ions is only 0.74 of the number in model A. Hence, the average density of electrons in model B is lower than in model A. The low electron density decreases the recombination rate of $Ne^{2+}$ so that the number of the $Ne^+$ ions is only 0.38 of the number of the $Ne^{2+}$ ions in model B. In the tail, for the same reason, the relative fraction of $Ne^+$ is generally lower than 0.3 in model B while that is always higher than 0.6 there in model A. The difference in the numbers between the $Ne^+$ and $Ne^{2+}$ ions is easily found in the right panel of Figure \ref{fig_modb} where the density distributions of $Ne^+$ and $Ne^{2+}$ ions are presented. The velocity field of photoionized gas in model B is presented in the right panel of Figure \ref{fig_discus}. In the boundary region, the axial components of the velocities are low, and the ionized materials in the rest part of the H II region all have an apparently blue-shifted velocity for the inclination of $0^o$.

\begin{figure}[!htp]
\centering
\includegraphics[scale=.33]{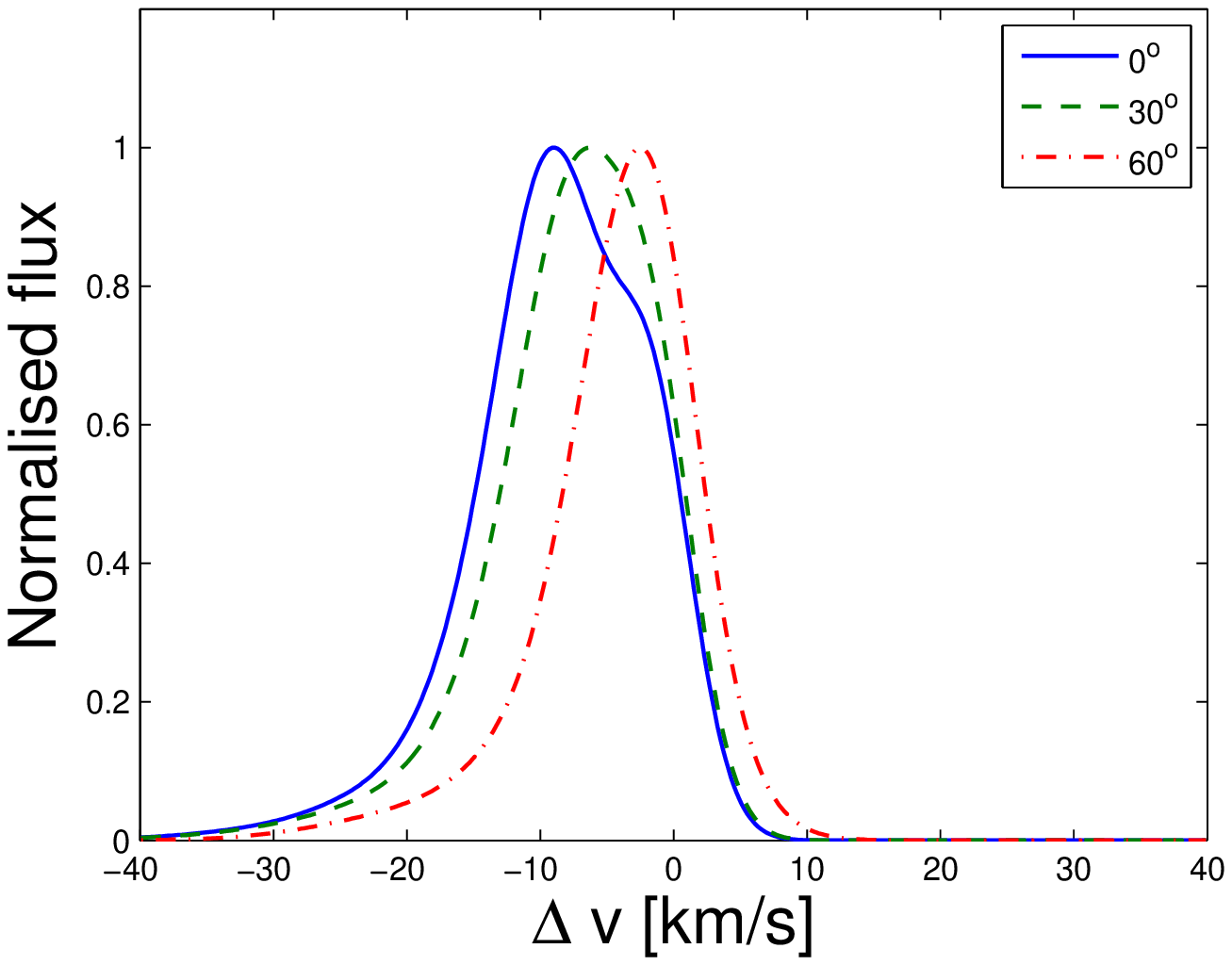}
\includegraphics[scale=.33]{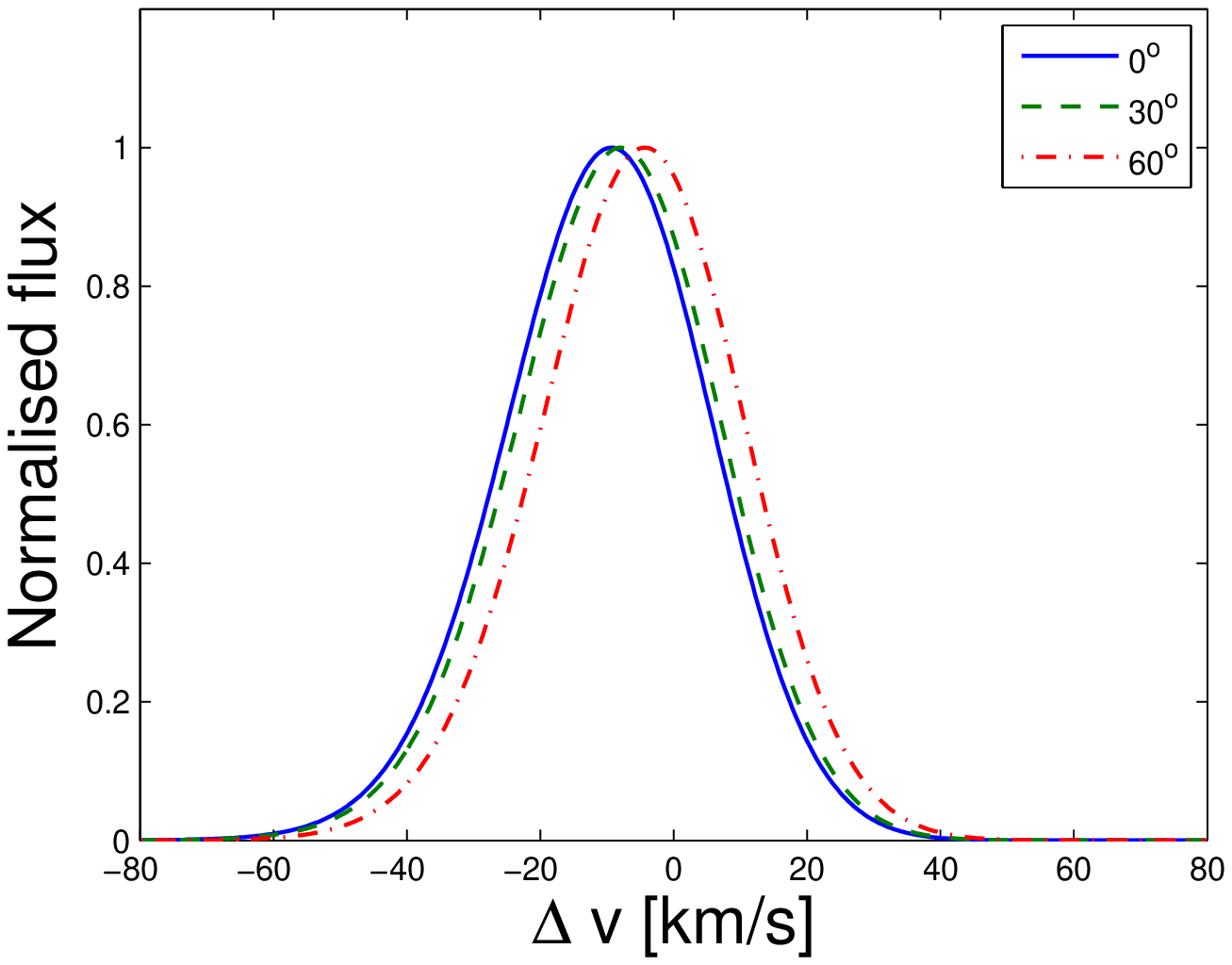}
\includegraphics[scale=.33]{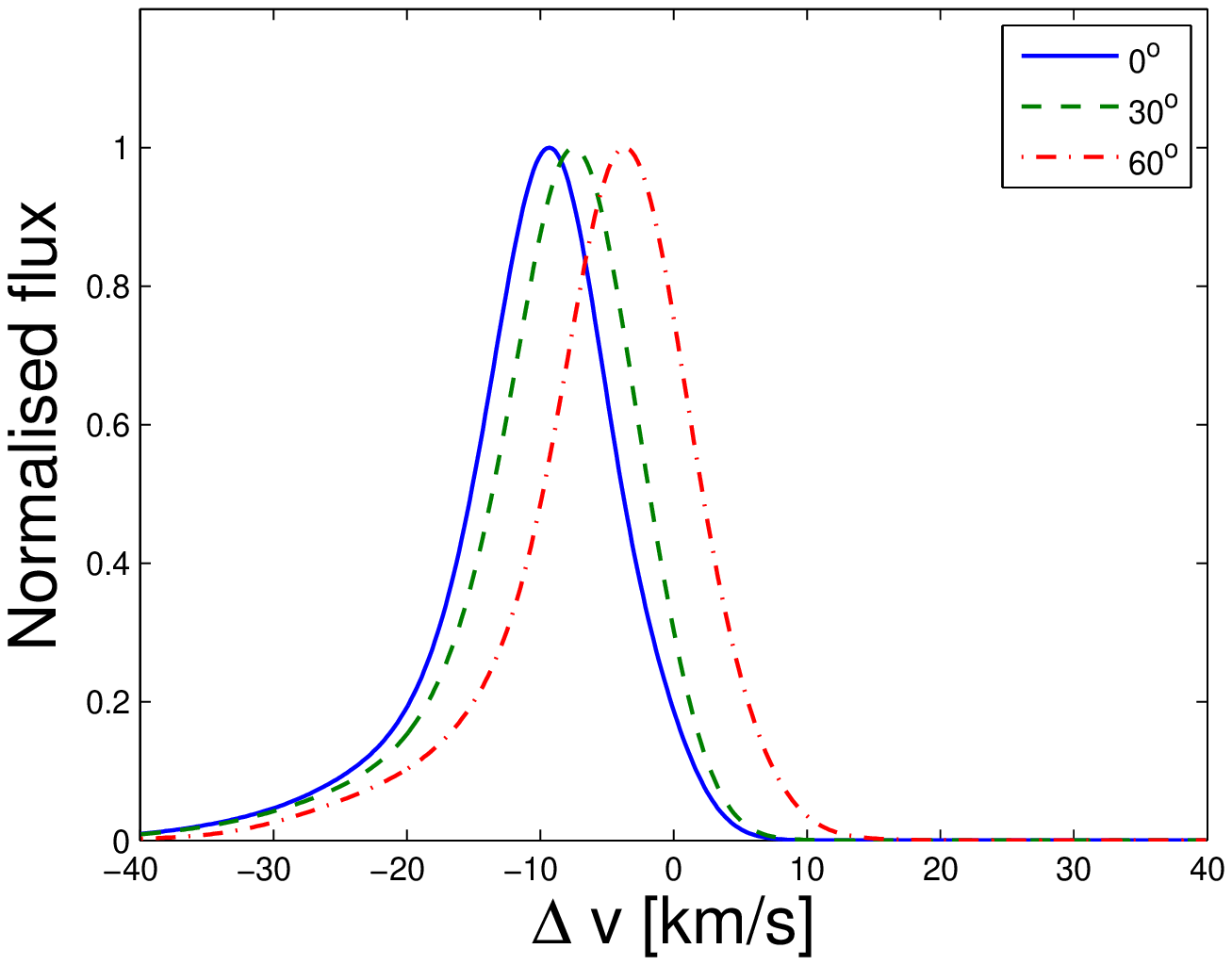}
\caption{profiles of the [Ne II] $12.81 \mu m$ line (left), the $H30\alpha$ line (middle) and the [Ne III] $15.55 \mu m$ line (right) from the cometary H II regions in model B for three inclination angles ($0^o,~30^o,~60^o$).}
\begin{flushleft}
\end{flushleft}
\label{fig_modbl}
\end{figure}

The line profiles from the H II region are plotted in Figure \ref{fig_modbl}. In model B, the line profiles are also asymmetrical as in model A. For the [Ne II] line, we separate the boundary region from the rest part of the H II region because the axial velocity gradually change along the symmetrical axis from $\sim0~km~s^{-1}$ at the boundary to $\sim-20~km~s^{-1}$ at the tail. The line profiles from these two regions are plotted in the right panel of Figure \ref{fig_modals}. The [Ne II] line profile is consist of two components: a narrow component with a slightly blue-shifted peak location contributed by the ionized gas in the boundary region and a highly blue-shifted broad component with the peak location at $v=-9.7~km~s^{-1}$ contributed by gas from the rest part. This suggests that gas in boundary region is not accelerated too much. Meanwhile, the ionized gas in the rest part has been accelerated to high velocities toward the tail direction. When the inclination angle increases, the highly blue-shifted component is affected strongly and the center of the component moves toward the red but a little change happens to the other component. It is obvious that the profiles of the [Ne II] line in model B are more biased to the left side than those in model A. In addition, because the number density is smaller than the critical densities of [Ne II] and [Ne III] lines in the H II region, the collisional de-excitation is not important for these lines. The fluxes of lines from a unit volume is $f_{ul}(i)\propto n_e^2$.  So the number of the [Ne II] line photons emitted from the dense boundary region is not negligible relative to the total number although the number of the $Ne^+$ in the boundary region is just 0.08 of the total number. For the $H30\alpha$ line and the [Ne III] line, the proportions of $H^+$ and $Ne^{2+}$ ions in the boundary region are smaller. So the proportions of the emission from boundary region in the total emission are both less than that for the [Ne II] line.

In Table \ref{tab_modb}, the peak locations, the FWHMs and the FWCVs in model B are provided. For the [Ne II] line profiles, the peak locations ($-8.9,~-6.3,~-2.5~km~s^{-1}$) and the FWCVs ($-8.52,~-7.38,~-4.28~km~s^{-1}$) are all blue-shifted. The values of the peak locations and the FWCVs in model B become less blue-shifted with increasing inclination angles, and they are all much more blue-shifted than the corresponding values in model A. For the $H30\alpha$ recombination line, we find that both the peak locations ($-9.4,~-7.8,~-4.2~km~s^{-1}$) and the FWCVs ($-10.13,~-8.77,~-5.06~km~s^{-1}$) are much more blue-shifted than those in model A because the main part of the ionized gas in model B are forced to flow away from the molecular cloud due to the density gradient. For the [Ne III] line, both of the FWCVs ($-10.81,~-9.37,~-5.45~km~s^{-1}$) and the peak locations ($-9.3,~-7.5,~-3.7~km~s^{-1}$) of the profiles are easily distinguished from those in model A. This suggests the ionized gas in the inner part of the H II region has a high velocity ($\sim-10~km~s^{-1}$) towards the tail. Also in Table \ref{tab_modb}, the [Ne III] line profiles have the most blue-shifted peak locations and FWCVs in the three lines. We find that the three line luminosities are lower in model B than in model A. This is due to the lower density in model B. 

\subsection{Comparison between Champagne Flow and Bow Shock Models}

We have calculated model A as a bow shock model and model B as a champagne flow model and obtain the line profiles, peak locations and FWCVs of the [Ne II] $12.81\mu m$ line, the $H30\alpha$ line and the [Ne III] $15.55\mu m$ line. It is worth to point out that line profiles and peak locations can be influenced by broadening. For example, if we change the broadening to be $5~km~s^{-1}$ , the peak location of the [Ne II] line profile for the inclination of $0^o$ in model A will be changed to be $1.1~km~s^{-1}$ from $3.1~km~s^{-1}$. But the FWCVs are independent of the broadening.

For the $H30\alpha$ line, both of the peak locations and the FWCVs in model A are much less blue-shifted than in model B for every inclinations. The FWCVs in model A are all higher than $-3.3~km~s^{-1}$, but those in model B are all lower than $-5~km~s^{-1}$. For the [Ne III] line, the difference is more apparent. The peak locations in model A are red-shifted while those in model B are blue-shifted for every inclinations. The FWCVs of the [Ne III] line in model A are all higher than $-3.5~km~s^{-1}$, but in model B they are all lower than $-5.4~km~s^{-1}$. Especially for the inclination of zero, the difference of the FWCVs in these two models is the largest and equal to $7.41~km~s^{-1}$. For the [Ne II] line, it is also the case that the emission is generally more blue-shifted in model B than in model A. And the FWCVs and the line profiles between these two models are also much different.

So it is possible to distinguish a champagne flow from a bow shock by using the [Ne II] line. And the difference between the champagne flow and bow shock is more obvious in the $H30\alpha$ line and the [Ne III] line. In order to check whether these conclusions are applicable in more general cases of champagne flows and bow shocks, we compute other models and test the effects of the density gradient, the velocity of the moving star, the mass of the star on the line profiles of the three lines.


\subsubsection{Density Gradient in the Champagne Flow}

A Champagne flow model with a shallow density gradient ($H=0.15pc$) is computed in model C. The other parameters are kept same as in model B. We cease the evolution in model C at $160,000yr$ as in model B. The line profiles are also roughly stable, and the pressure equilibrium has been formed.

\begin{figure}[!htp]
\centering
\includegraphics[scale=.33]{figmodbll1.eps}
\includegraphics[scale=.33]{figmodbll2.eps}
\includegraphics[scale=.33]{figmodbll3.eps}
\includegraphics[scale=.33]{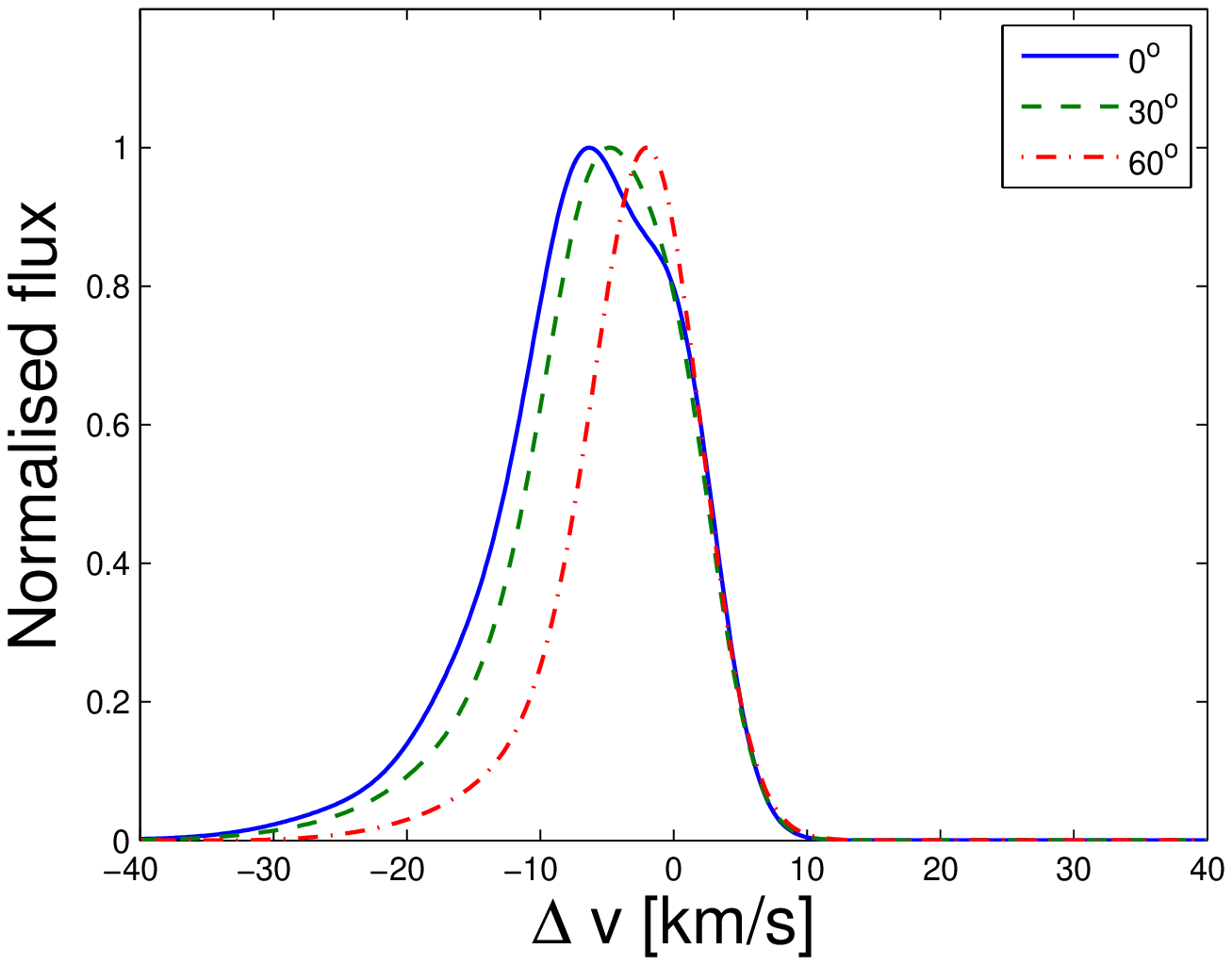}
\includegraphics[scale=.33]{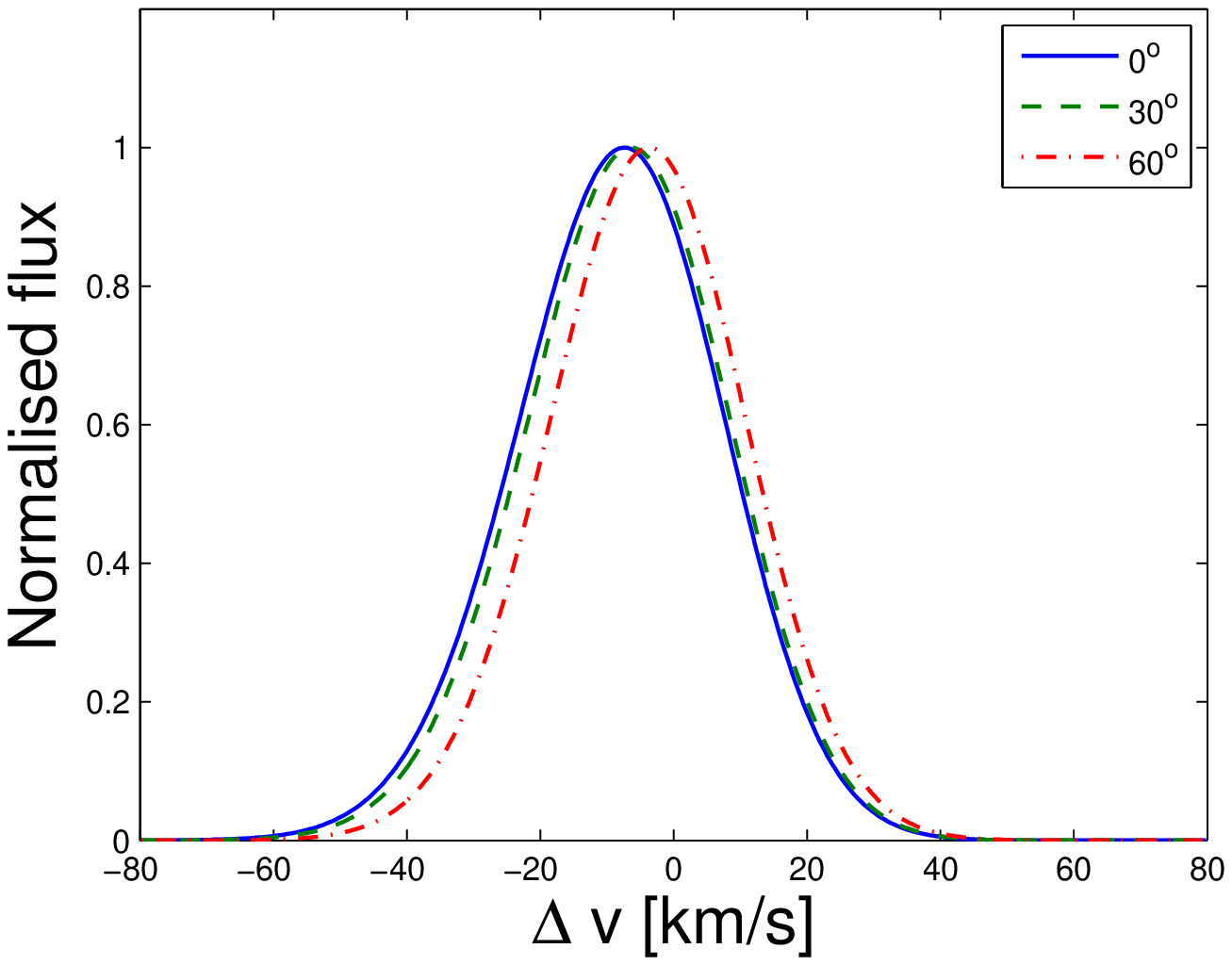}
\includegraphics[scale=.33]{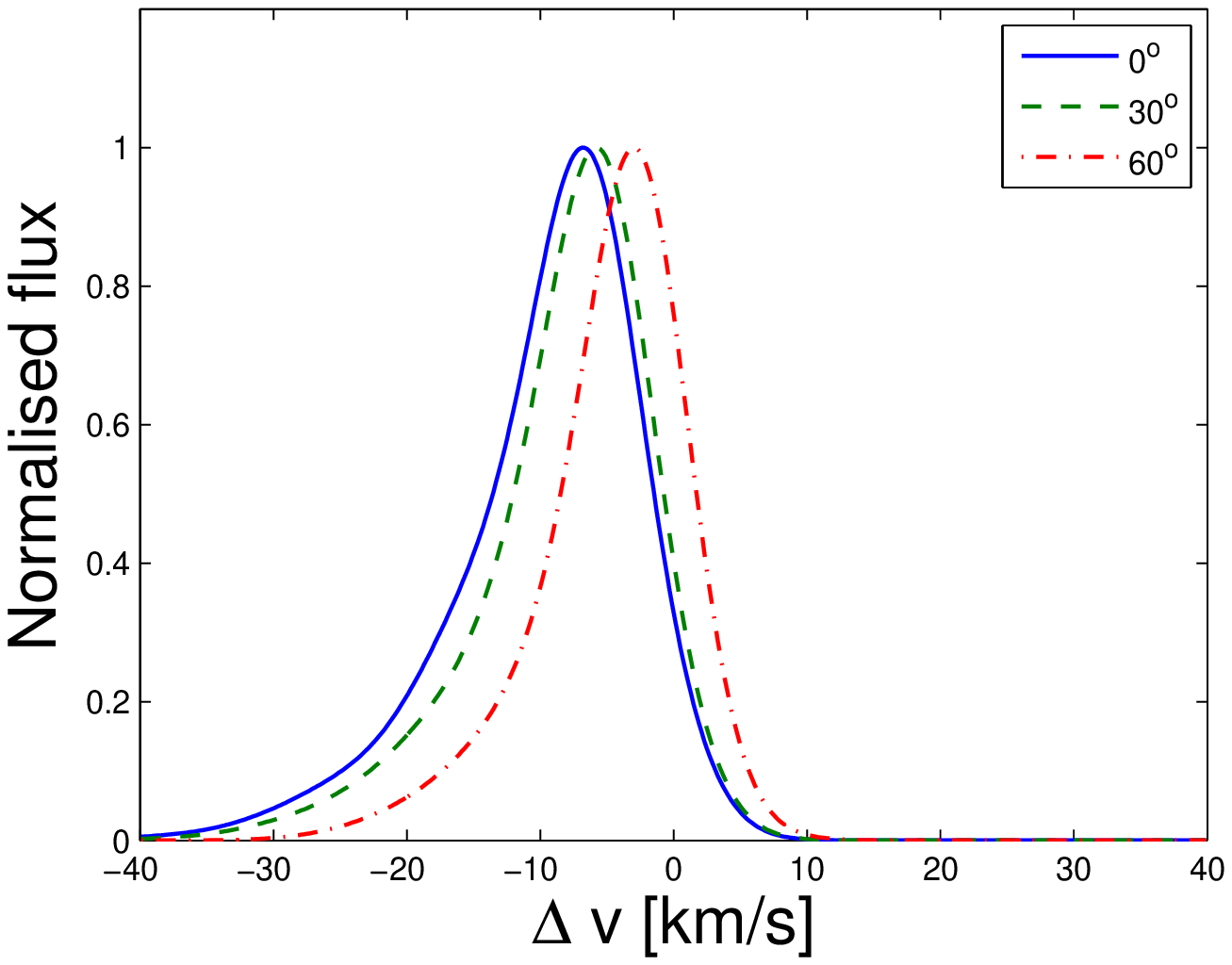}
\caption{The profiles of the [Ne II] $12.81 \mu m$ line (left panels), the $H30\alpha$ line (middle panels) and the [Ne III] $15.55 \mu m$ line (right panels) from the H II regions for three inclination angles in model B (top panels) and model C (bottom panels).}
\begin{flushleft}
\end{flushleft}
\label{fig_modcde}
\end{figure}

The line profiles and properties of the H II regions for model B and C are shown in Figure \ref{fig_modcde} and Table \ref{tab_modcde}.
In model C, the H II regions can also be divided into a boundary region and the rest region as in model B. The velocities in the boundary region are only slightly blue-shifted and close to zero, and the gas in the rest part of the H II region has a obviously blue-shifted velocity. Because of the shallow density gradient, the peak location of the [Ne II] line profile from the rest part of the H II region is at $-7.2~km~s^{-1}$. This suggests that acceleration of the ionized gas is smaller in model C than in model B. Beside this, the line profiles for model C are similar to those for model B.

In Table \ref{tab_modcde}, we find that the peak locations and the FWCVs of the three lines are all blue-shifted. The values of the FWCVs and the peak locations in model C are slightly less blue-shifted than the corresponding values in model B. And as in model B, the FWCVs and the peak locations of the [Ne III] line have the most blue-shifted values in the line profiles of the three lines.

 Our calculations suggest that the line profiles become more blue-shifted with the increasing density gradient for the champagne flow models.

\begin{table}[!htp]\footnotesize
\centering
\begin{tabular}{l|llll|l}
\hline
Model B $H=0.05pc$  \\
\hline
Line & Inclination & Peak & FWHM & FWCV & Luminosity \\
 & & ($km~s^{-1}$) & ($km~s^{-1}$) & ($km~s^{-1}$) & ($erg~s^{-1}$) \\
\hline
[Ne II] $12.81 \mu m$ & $0^o$ & -8.9 & 15.4 & -8.52 & $3.43\times10^{35}$ \\
 & $30^o$ & -6.3 & 13.8 & -7.38 \\
 & $60^o$ & -2.5 & 10.8 & -4.28 \\
\hline
$H30\alpha$ & $0^o$ & -9.4 & 36.0 & -10.13 & $6.08\times10^{29}$ \\
 & $30^o$ & -7.8 & 35.6 & -8.77 \\
 & $60^o$ & -4.2 & 35.6 & -5.06 \\
\hline
[Ne III] $15.55 \mu m$ & $0^o$ & -9.3 & 11.4 & -10.81 & $6.60\times10^{35}$ \\
 & $30^o$ & -7.5 & 11.8 & -9.37 \\
 & $60^o$ & -3.7 & 12.0 & -5.45 \\
\hline
Model E $H=0.15pc$   \\
\hline
Line & Inclination & Peak & FWHM & FWCV & Luminosity \\
 & & ($km~s^{-1}$) & ($km~s^{-1}$) & ($km~s^{-1}$) & ($erg~s^{-1}$) \\
\hline
[Ne II] $12.81 \mu m$ & $0^o$ & -6.3 & 15.6 & -6.60 & $3.68\times10^{35}$\\
 & $30^o$ & -4.7 & 13.6 & -5.70 \\
 & $60^o$ & -2.1 & 9.8 & -3.29 \\
\hline
$H30\alpha$ & $0^o$ & -7.4 & 36.4 & -8.32 & $4.90\times10^{29}$\\
 & $30^o$ & -6.2 & 36.0 & -7.21 \\
 & $60^o$ & -3.8 & 34.4 & -4.16 \\
\hline
[Ne III] $15.55 \mu m$ & $0^o$ & -6.7 & 12.2 & -9.48 & $4.70\times10^{35}$ \\
 & $30^o$ & -5.9 & 11.2 & -8.24 \\
 & $60^o$ & -2.9 & 10.2 & -4.77 \\
\hline
\end{tabular}
\caption{Peaks, FWHMs and flux weighted central velocities(FWCV) of lines at different angles in model B and C.\label{tab_modb} \label{tab_modcde}}
\end{table}

\subsubsection{Stellar Motion in the Bow Shock}
\label{sect:stellar motion}

\begin{figure}[!htp]
\centering
\includegraphics[scale=.33]{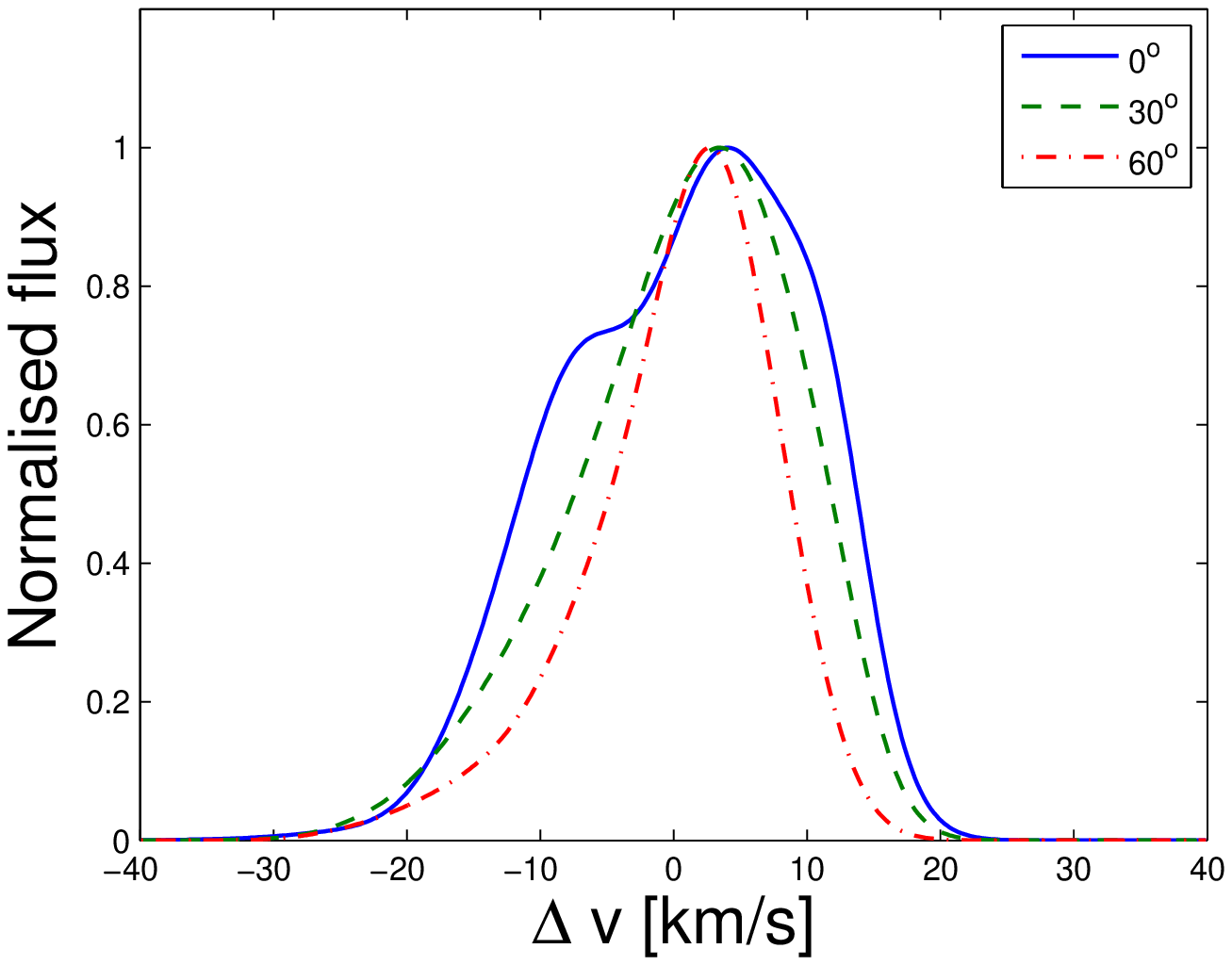}
\includegraphics[scale=.33]{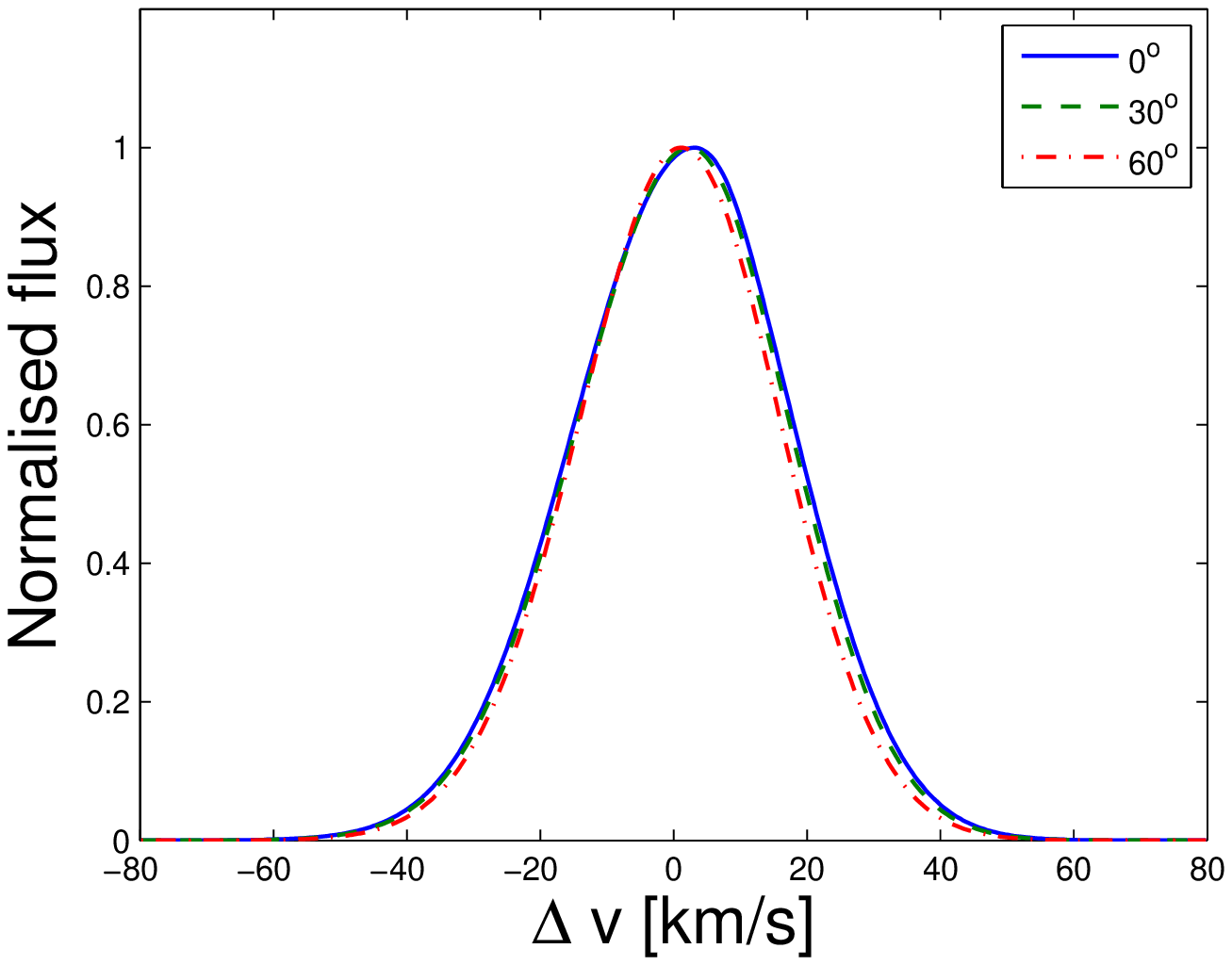}
\includegraphics[scale=.33]{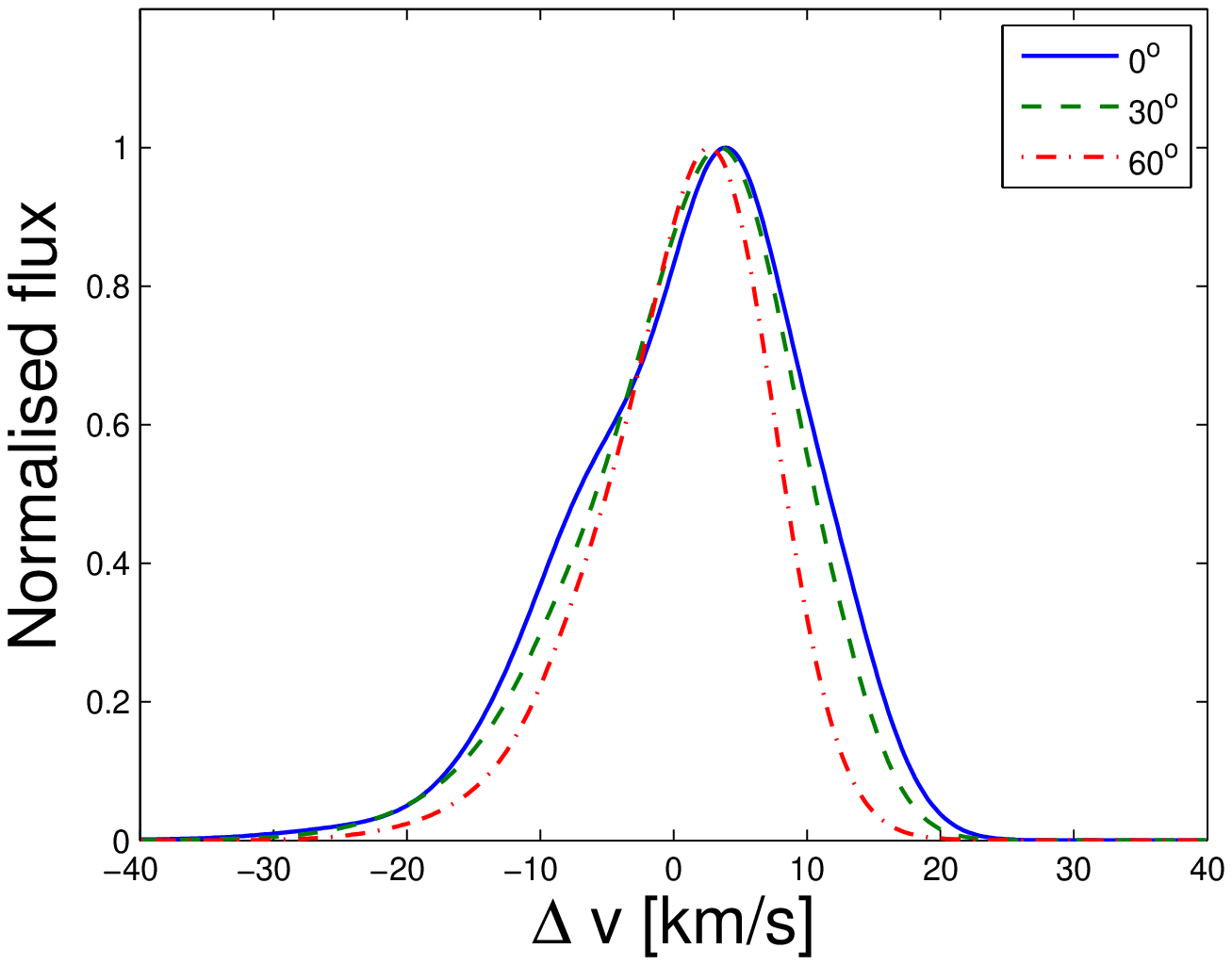}
\includegraphics[scale=.33]{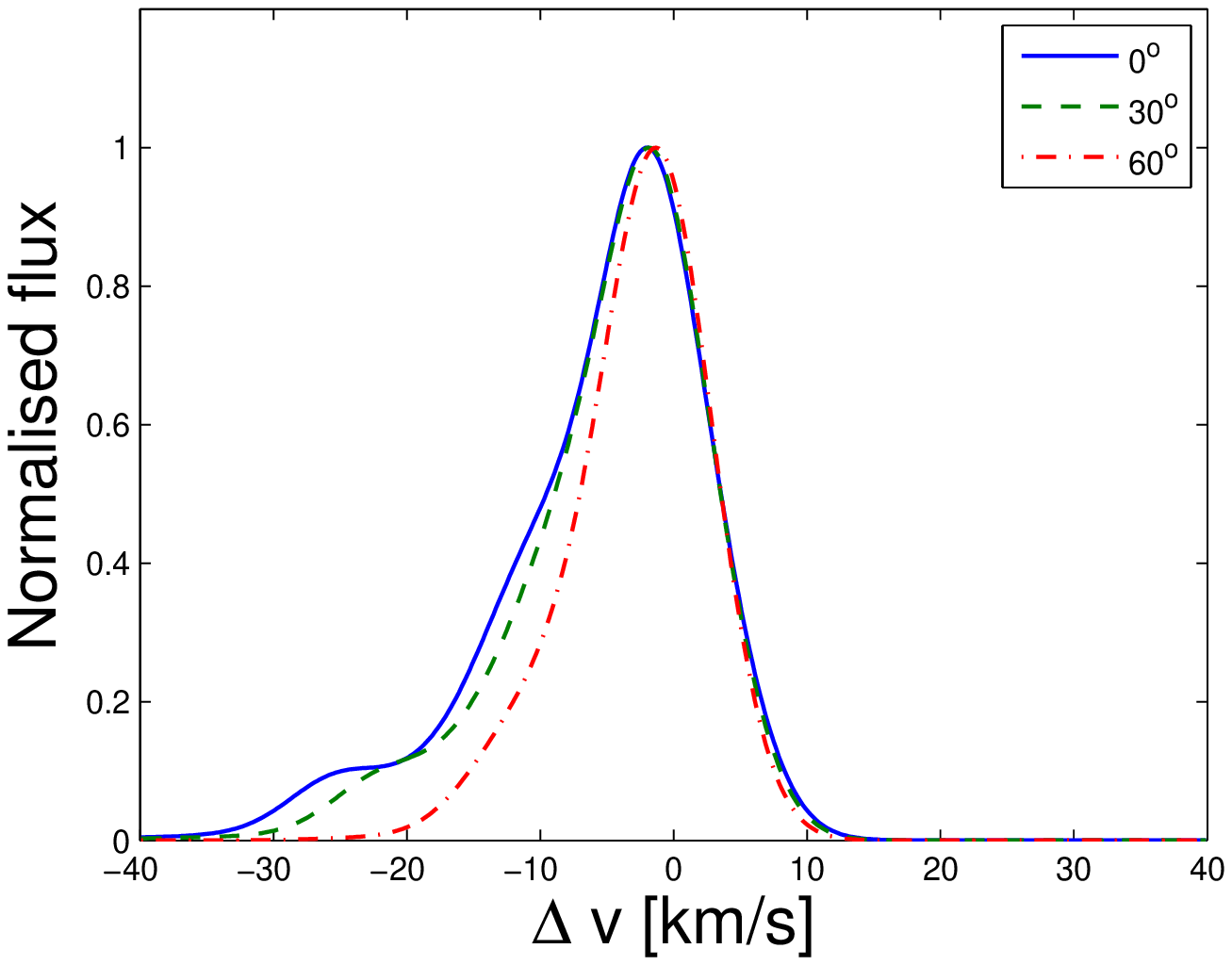}
\includegraphics[scale=.33]{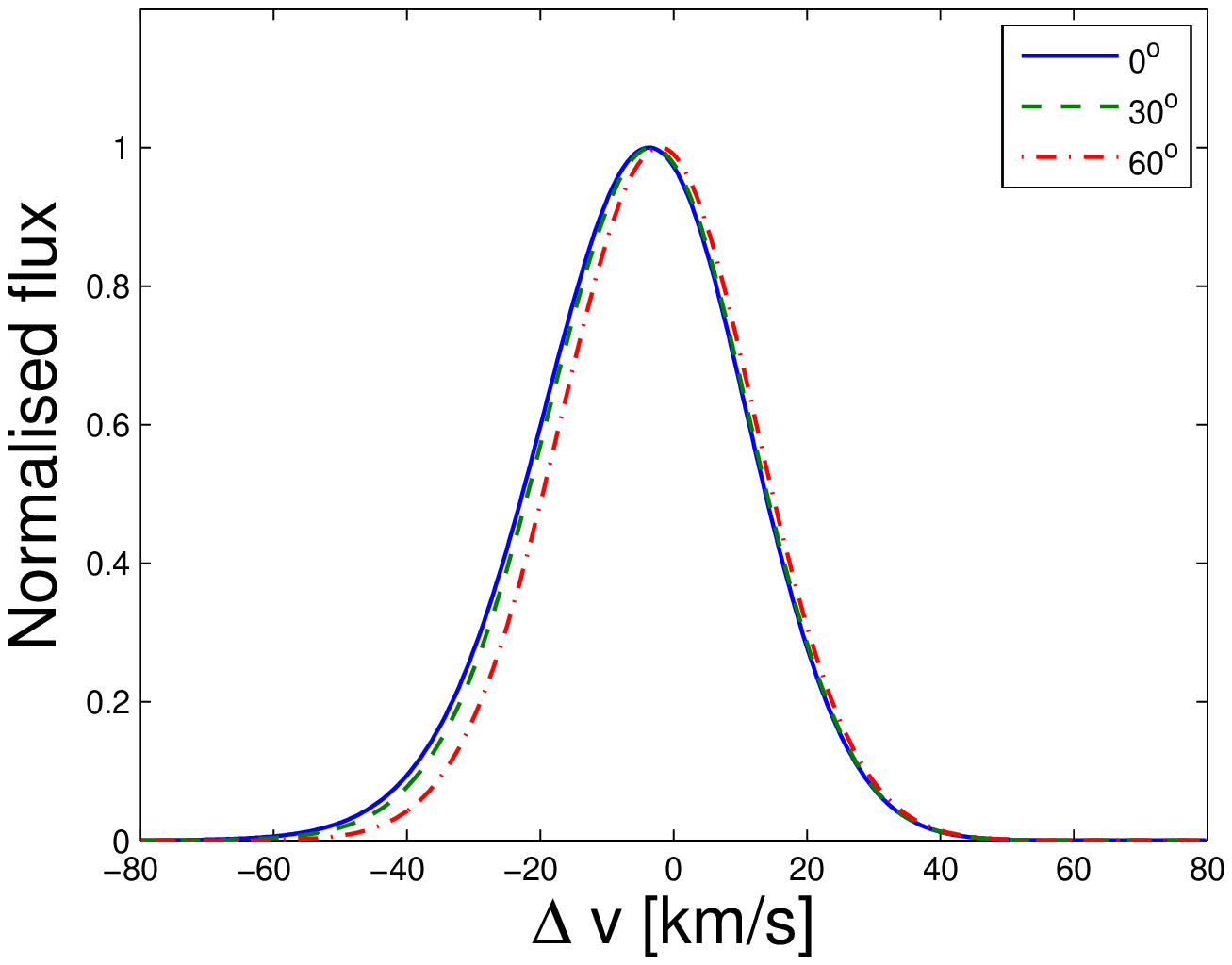}
\includegraphics[scale=.33]{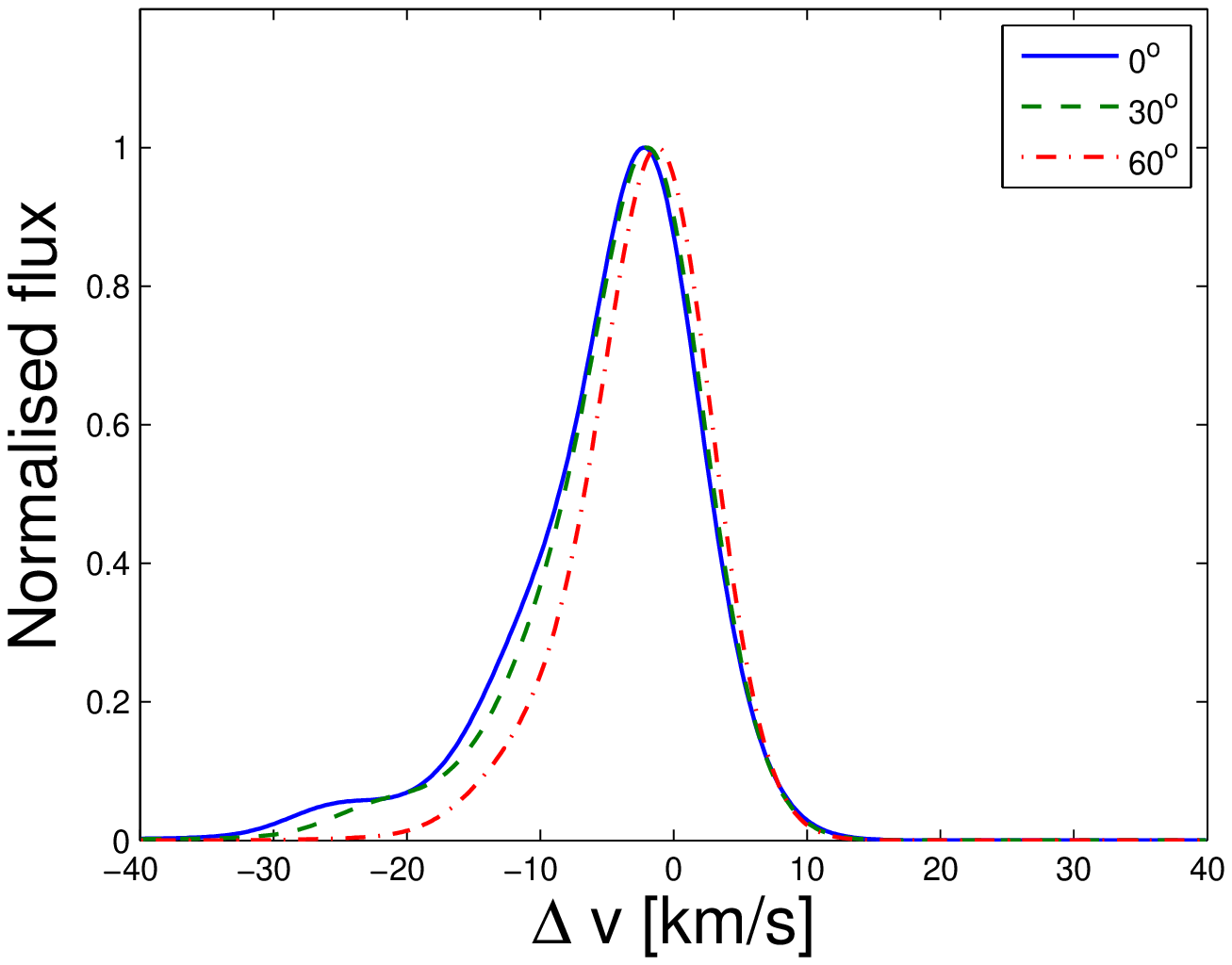}
\caption{Same as Figure\ref{fig_moda}, but for model D(top panels) and model E(bottom panels). There are the profiles of the [Ne II] $12.81 \mu m$ line (left panels), the $H30\alpha$ line (middle panels) and the [Ne III] $15.55 \mu m$ line (right panels) from the cometary H II regions in model A for three inclination angles ($0^0,~30^0,~60^0$).}
\begin{flushleft}
\end{flushleft}
\label{fig_modfg}
\end{figure}

The difference among model A, D, and E is the stellar velocity with respect to the ambient molecular clouds ($v_\ast=10,~15~\textrm{and}~5~km~s^{-1}$).
The line profiles are plotted in Figure \ref{fig_modfg}. The peak locations and the FWHMs are shown in the Table \ref{tab modfg}. We find that the line profiles are more biased towards the red-shifted side with the increasing stellar velocity. The peak locations and the FWCVs of the three line profiles in model D are all red-shifted. On the contrary, these values in model E are all blue-shifted. This is because the proportion of ionized gas with a red-shifted velocity is the highest in model D and is the lowest in model E. As in model A, the red-shifted materials are mainly in the boundary region and the head region. In model D, the [Ne II] line profile is also consist of the components from the red-shifted region and from the blue-shifted region. But this is not obvious in the profiles of the [Ne III] and the $H30\alpha$ line due to the different relative fraction of $Ne^{2+}$ and $H^+$ and the large broadening of $H^+$. In model E, because the proportion and the velocities of red-shifted ionized gas are both low, the contribution to the red-shifted part of the [Ne II] line profile is small as well. In our results of the simulations, the size of the H II region decreases with the increasing stellar velocity. A shorter distance due to the smaller size causes the flux of the ionizing photons to be stronger so that the ionization rate from $Ne^+$ to $Ne^{2+}$ rises. But the density of electrons could be higher and increase the recombination rate of $Ne^{2+}$ ions. The comparison of the line luminosities of the [Ne II] line and the [Ne III] line shows that higher stellar velocity causes lower ratio of the [Ne II] line luminosity to the [Ne III] line luminosity. This suggests that the increase of the ionization rate from $Ne^+$ to $Ne^{2+}$ is more than that of the recombination rate.

In bow shock models, the ionized gas compresses the neutral and cold materials ahead of the moving star into a dense shell while the density of the ionized gas in H II region is much lower than the density of the shell. When the evolution reaches quasi-steady state, the ionization front and the shock front are motionless relative to the star. Along the arched shell from the apex to the tail, the axial velocity of the neutral gas decreases from the stellar velocity to zero. In the ionized region, the gases just peeled from the shell by the ionization are dense ($\sim 10000~cm^{-3}$) and have a velocity slightly less than the velocity of nearby neutral gases in the shell. With the expansion and moving to the tail of the H II region, the densities of the ionized gases gradually decrease to lower values ($\sim 50~cm^{-3}$). This leads to a pressure gradient from the head to the tail as in a champagne flow model. Although the ionized gases ahead of the star are accelerated by the stellar wind, the advection in the head region pushes these gases to the sides where the pressure gradient is dominant. So if the stellar velocity are higher, it need more time for the pressure gradient to accelerate the ionized gas from shell toward the blue-shifted direction, so that the proportion of the red-shifted ionized gas in the H II region will be higher. In model E, the effect of the pressure gradient is dominant in the H II region due to the low stellar velocity. Hence, the line profiles for model E in Figure \ref{fig_modfg} are more biased to the left side than those in model A and model D. And the peak locations and the FWCVs in model E are not easily distinguished from those in model C.



\begin{table}[!htp]\footnotesize
\centering
\begin{tabular}{l|llll|l}
\hline
Model D $v_\ast=15km~s^{-1}$  \\
\hline
Line & Inclination & Peak & FWHM & FWCV & Luminosity \\
 & & ($km~s^{-1}$) & ($km~s^{-1}$) & ($km~s^{-1}$) & ($erg~s^{-1}$) \\
\hline
[Ne II] $12.81 \mu m$ & $0^o$ & 4.1 & 25.2 & 0.73 & $7.79\times10^{35}$ \\
 & $30^o$ & 3.5 & 19.2 & 0.63 \\
 & $60^o$ & 2.7 & 13.6 & 0.36 \\
\hline
$H30\alpha$ & $0^o$ & 5.8 & 36.4 & 1.14 & $1.18\times10^{30}$ \\
 & $30^o$ & 3.8 & 35.6 & 0.98 \\
 & $60^o$ & 1.0 & 34.4 & 0.57 \\
\hline
[Ne III] $15.55 \mu m$ & $0^o$ & 3.9 & 18.8 & 1.26 & $1.14\times10^{36}$ \\
 & $30^o$ & 3.5 & 16.4 & 1.08 \\
 & $60^o$ & 2.7 & 13.4 & 0.62 \\
\hline
Model E $v_\ast=5km~s^{-1}$  \\
\hline
Line & Inclination & Peak & FWHM & FWCV & Luminosity \\
 & & ($km~s^{-1}$) & ($km~s^{-1}$) & ($km~s^{-1}$) & ($erg~s^{-1}$) \\
\hline
[Ne II] $12.81 \mu m$ & $0^o$ & -1.9 & 13.2 & -5.49 & $8.24\times10^{35}$\\
 & $30^o$ & -1.9 & 12.6 & -4.80 \\
 & $60^o$ & -1.3 & 10.6 & -2.80 \\
\hline
$H30\alpha$ & $0^o$ & -3.8 & 36.4 & -5.09 & $9.54\times10^{29}$ \\
 & $30^o$ & -3.4 & 36.0 & -4.41 \\
 & $60^o$ & -2.2 & 34.4 & -2.55 \\
\hline
[Ne III] $15.55 \mu m$ & $0^o$ & -2.3 & 11.4 & -4.78 & $8.09\times10^{35}$ \\
 & $30^o$ & -1.9 & 11.2 & -4.17 \\
 & $60^o$ & -1.3 & 10.4 & -2.43 \\
\hline
\end{tabular}
\caption{Peaks, FWHMs and flux weighted central velocities (FWCV) of lines in model D and E.\label{tab modfg}}
\end{table}

\subsubsection{the Mass of the star}

We assume a less massive star ($M_*=21.9~M_\odot$) in model F and G. This leads to a low effective temperature $35000K$, a weak stellar wind and a weak radiation (See Table\ref{tab_modc}). The other parameters in model F and model G are kept same as in model B and model A, respectively. 

In model F, because of the weaker stellar wind and ionizing radiation, the size of the H II region is also smaller than in model B. The lower effective temperature causes the relative fraction of $Ne^+$ ions to be higher. For example, in the tail, $X(Ne^+)$ is generally higher than $0.6$ in model F but lower than $0.3$ in model B.

The H II region can also be divided into the boundary region and the rest region. For the [Ne II] line, the proportion of the photons emitted from the boundary region is lower than in model B. This is due to the higher relative fraction of $Ne^{+}$ in the inner part of the H II region and causes the lower normalized flux at the velocities close to $0$ in the [Ne II] line profile. The other two line profiles in model F are similar to those in model B.

In Table \ref{tab_modhi}, the properties of the line profiles are shown. The FWCVs in model F are more blue-shifted than the corresponding values in model B. This is related with the higher relative fraction of $Ne^+$ in the inner part and the lower fluxes emitted from the boundary region where the velocities close to zero. In addition, the low effective temperature and the ionization radiation of the star lead to similar distributions of $Ne^+$ and $Ne^{2+}$ ions in the H II region so that the FWCVs of the [Ne II] line, the $H30\alpha$ line and the [Ne III] line are closer to each others in model F than in model B. Because of the high relative fraction of $Ne^+$, the line luminosity of the [Ne II] line is higher than that of the [Ne III] line in model F.


The weaker EUV flux and the weaker ram-pressure of the stellar wind in model G result in a smaller H II region and result in a smaller mass ratio between the head region and the entire H II region than in model A ($0.11/0.15$ in model G/A). The proportion of the red-shifted ionized gas is also lower than in model A ($0.13/0.18$ in model G/A). The lower effective temperature leads to a higher relative fraction of $Ne^+$ ions in the H II region as in model F.

The peak locations, line luminosities and the flux weighted central velocities in model G are also presented in Table \ref{tab_modhi}. The values of the FWCVs of the three lines are close to each others for every inclination as in model F. But it is only a coincidence that the FWCVs of the [Ne II] line are approximately equal to those of the [Ne III] line. The FWCVs in model G are all more blue-shifted than the corresponding values in model A. This is attributed to the lower proportion of the red-shifted ionized gas mentioned above. It is also obvious that the line luminosity of the [Ne II] line presented is much stronger than that of the [Ne III] line in model G due to the low effective temperature.


\begin{figure}[!htp]
\centering
\includegraphics[scale=.33]{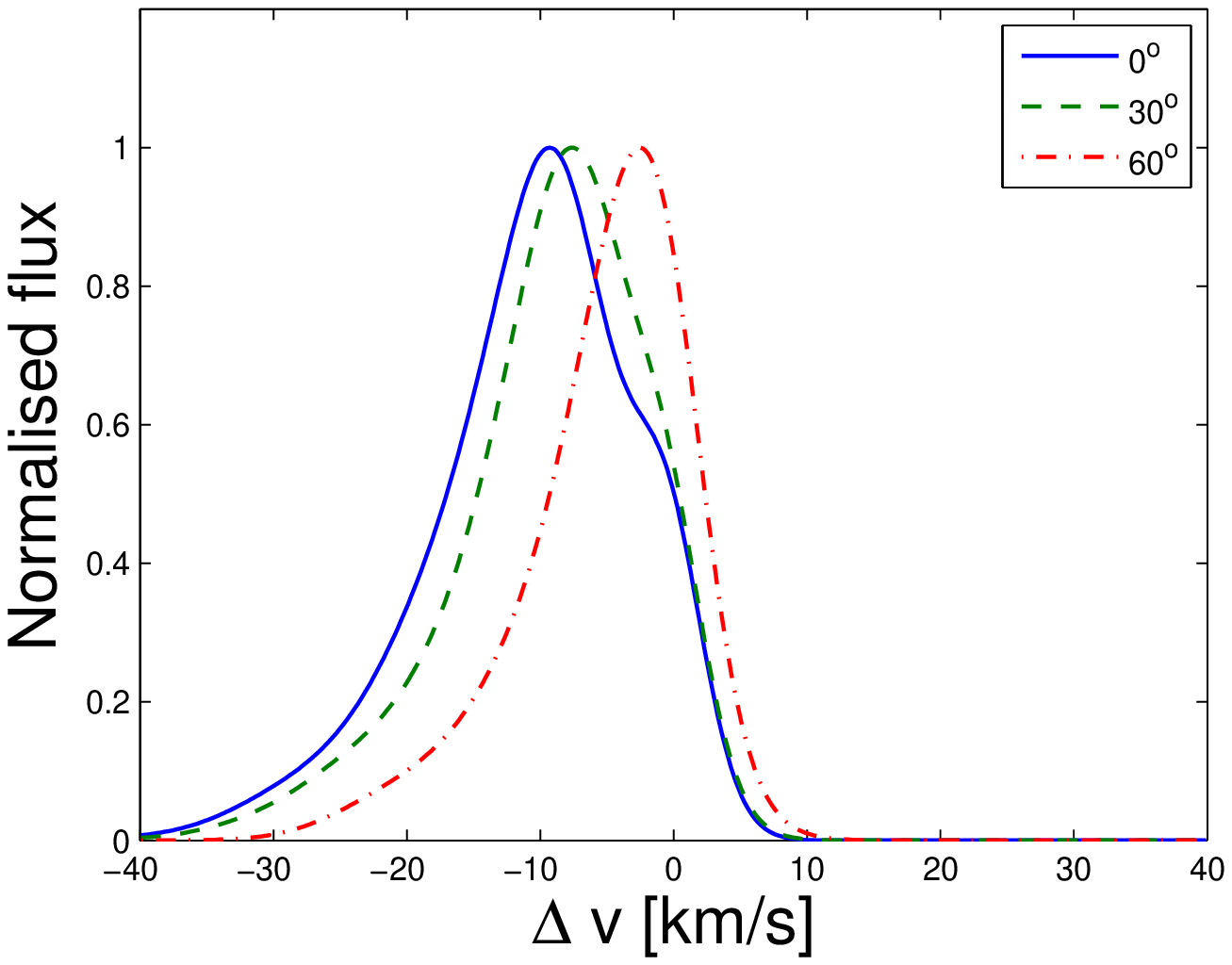}
\includegraphics[scale=.33]{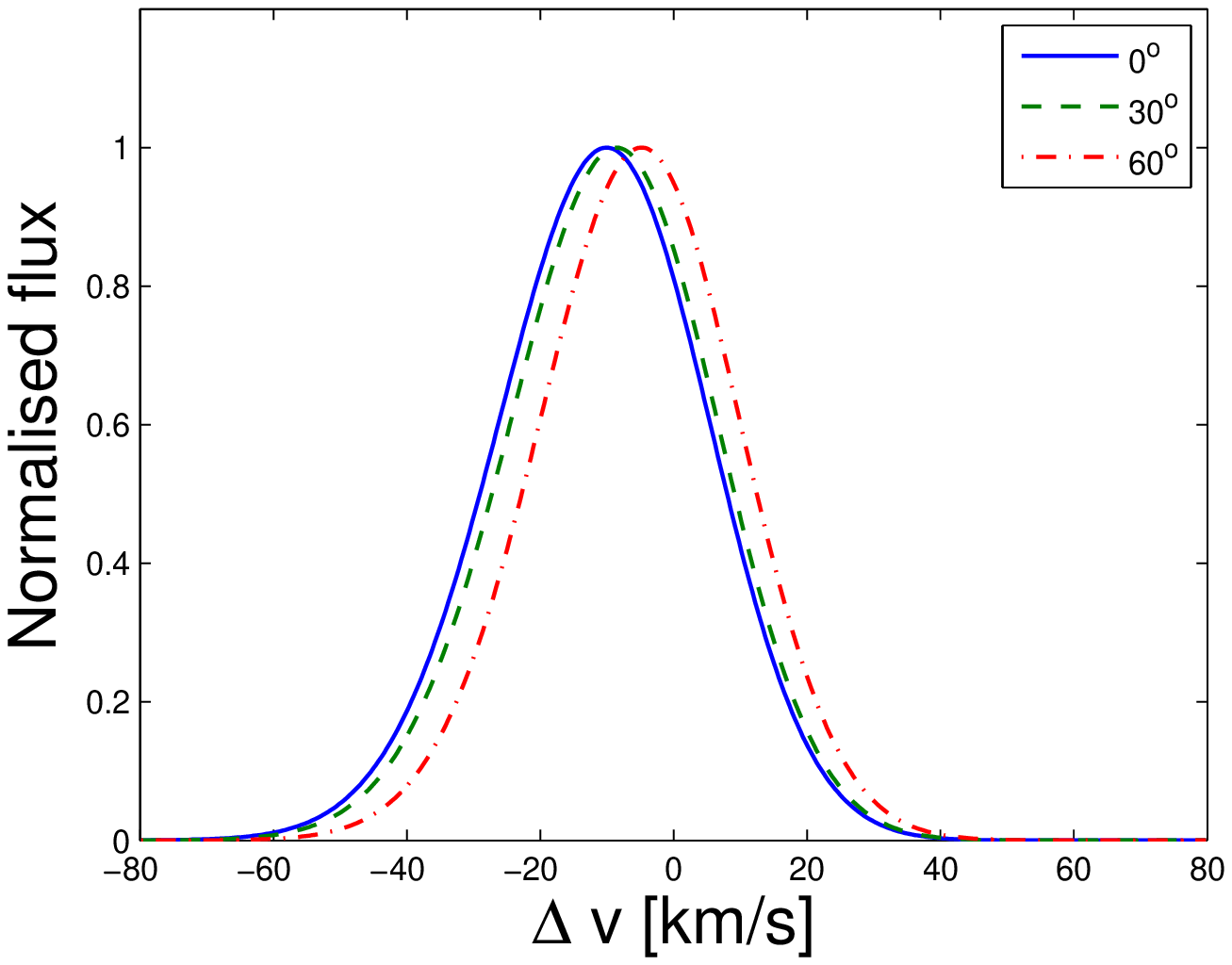}
\includegraphics[scale=.33]{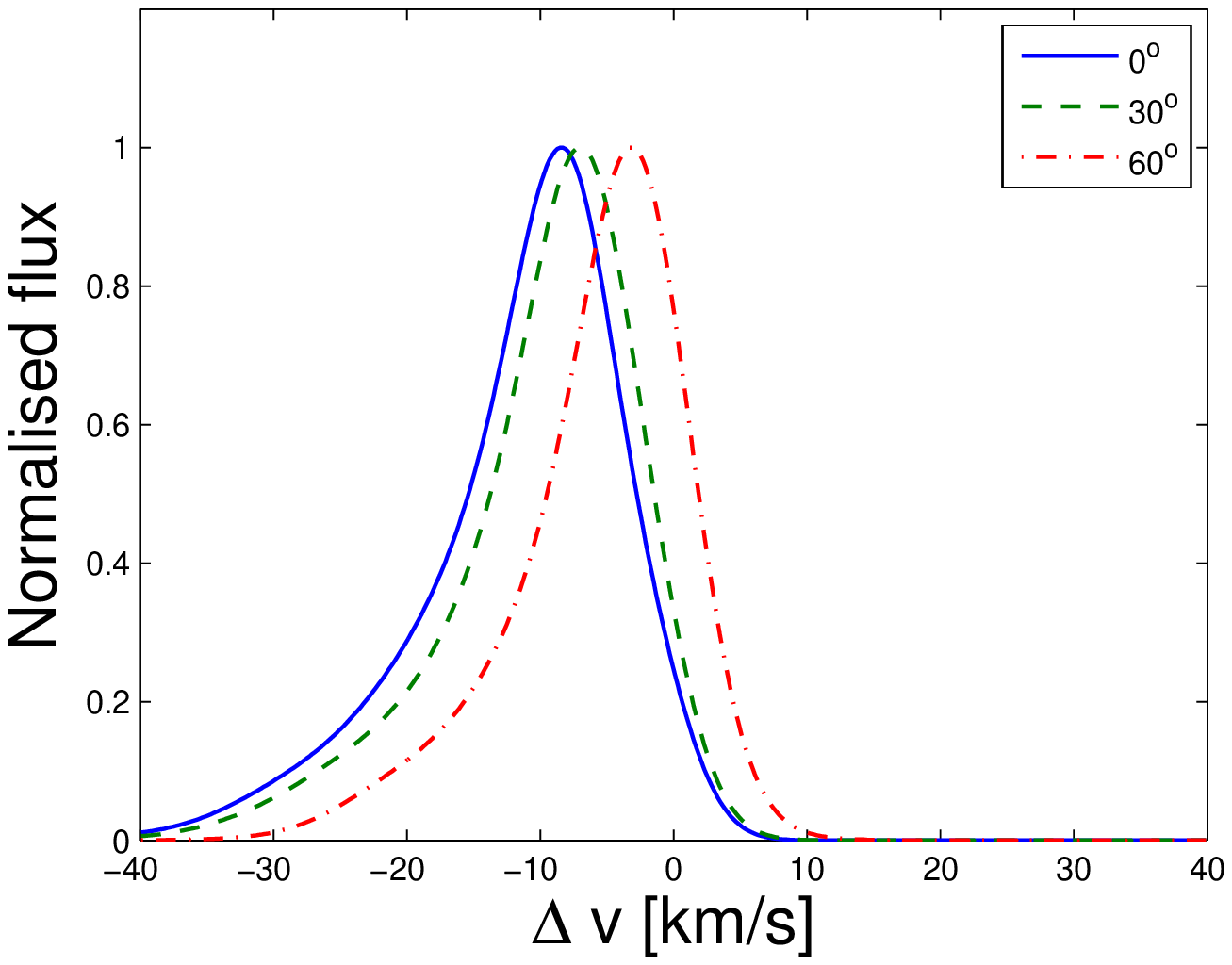}
\includegraphics[scale=.33]{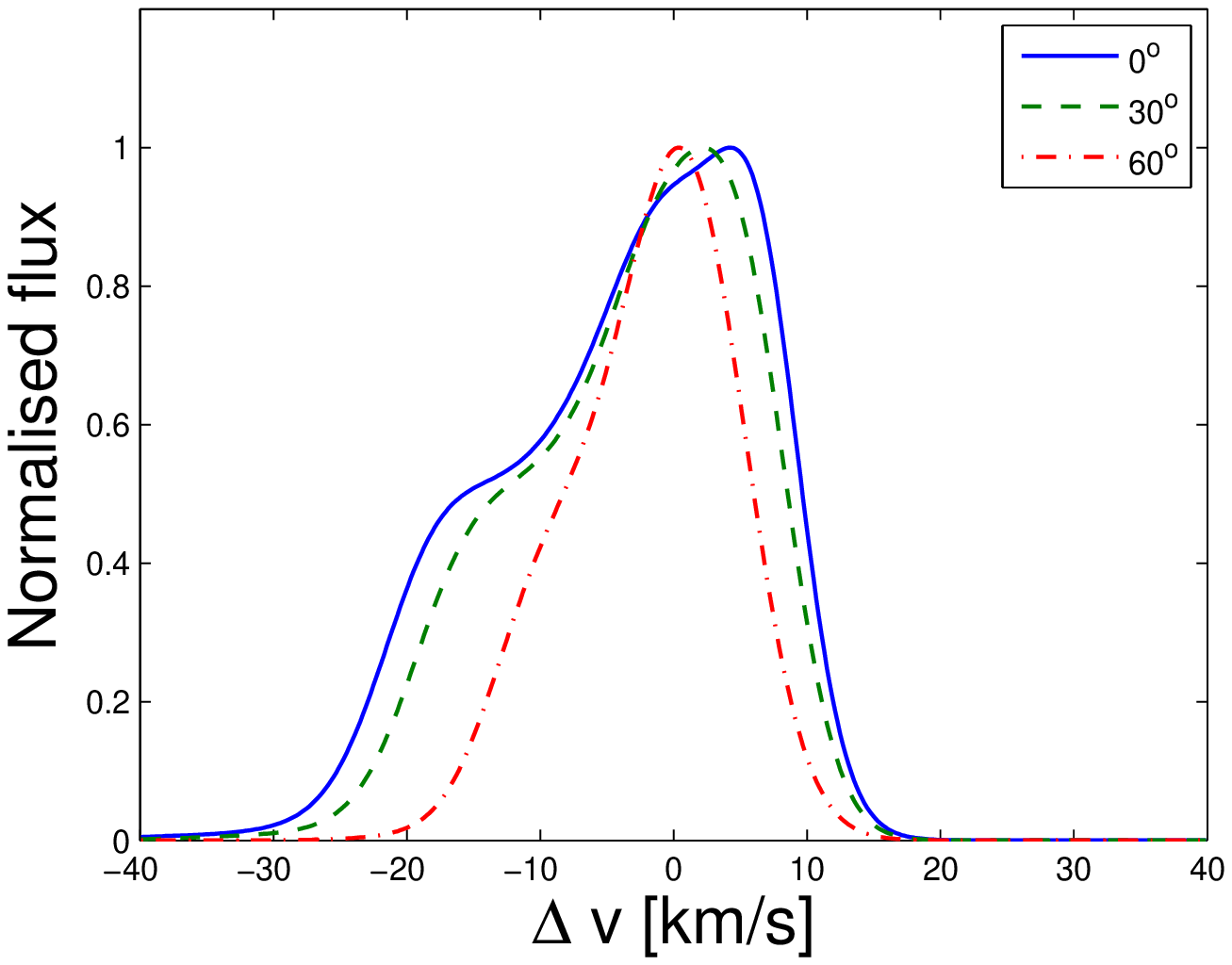}
\includegraphics[scale=.33]{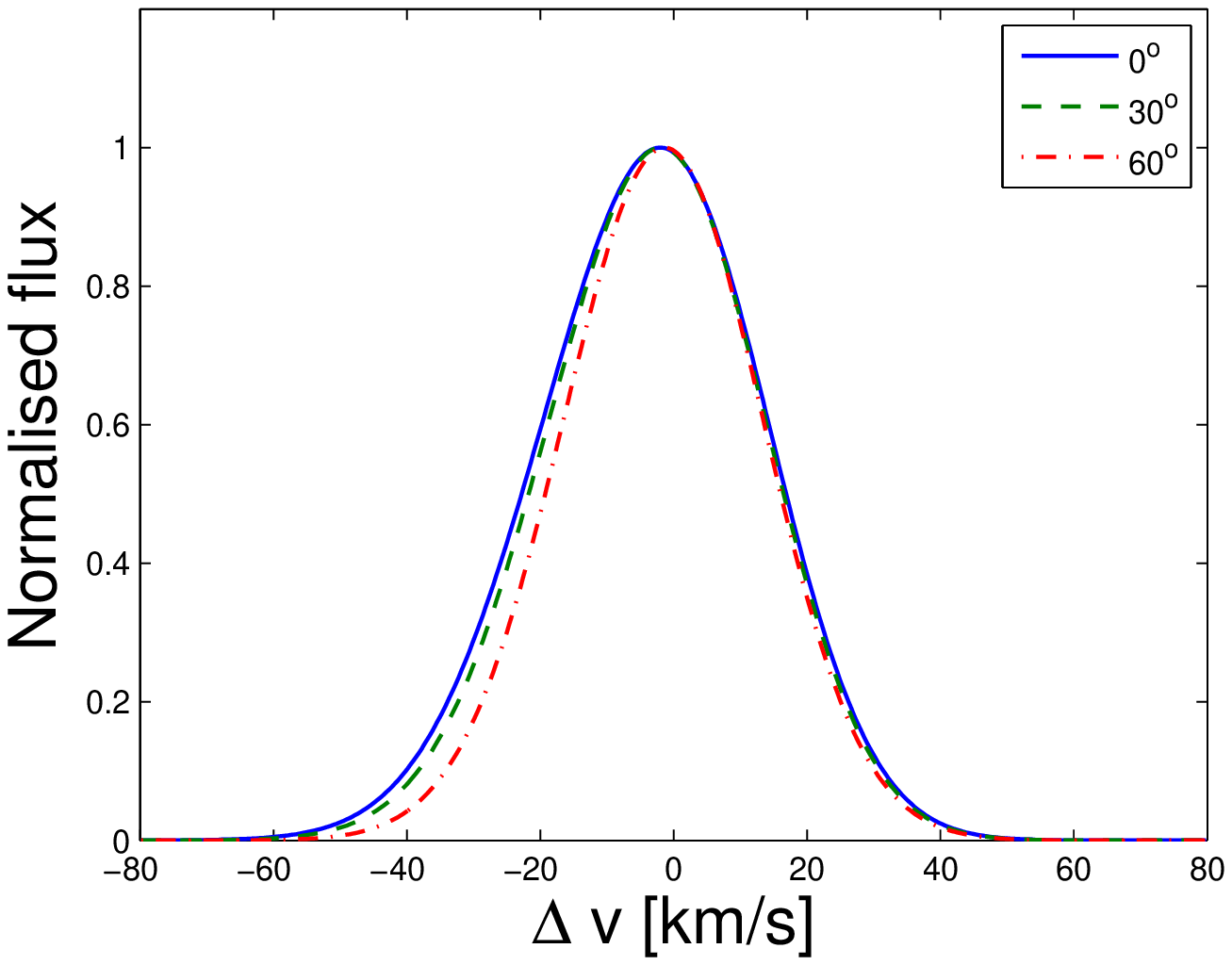}
\includegraphics[scale=.33]{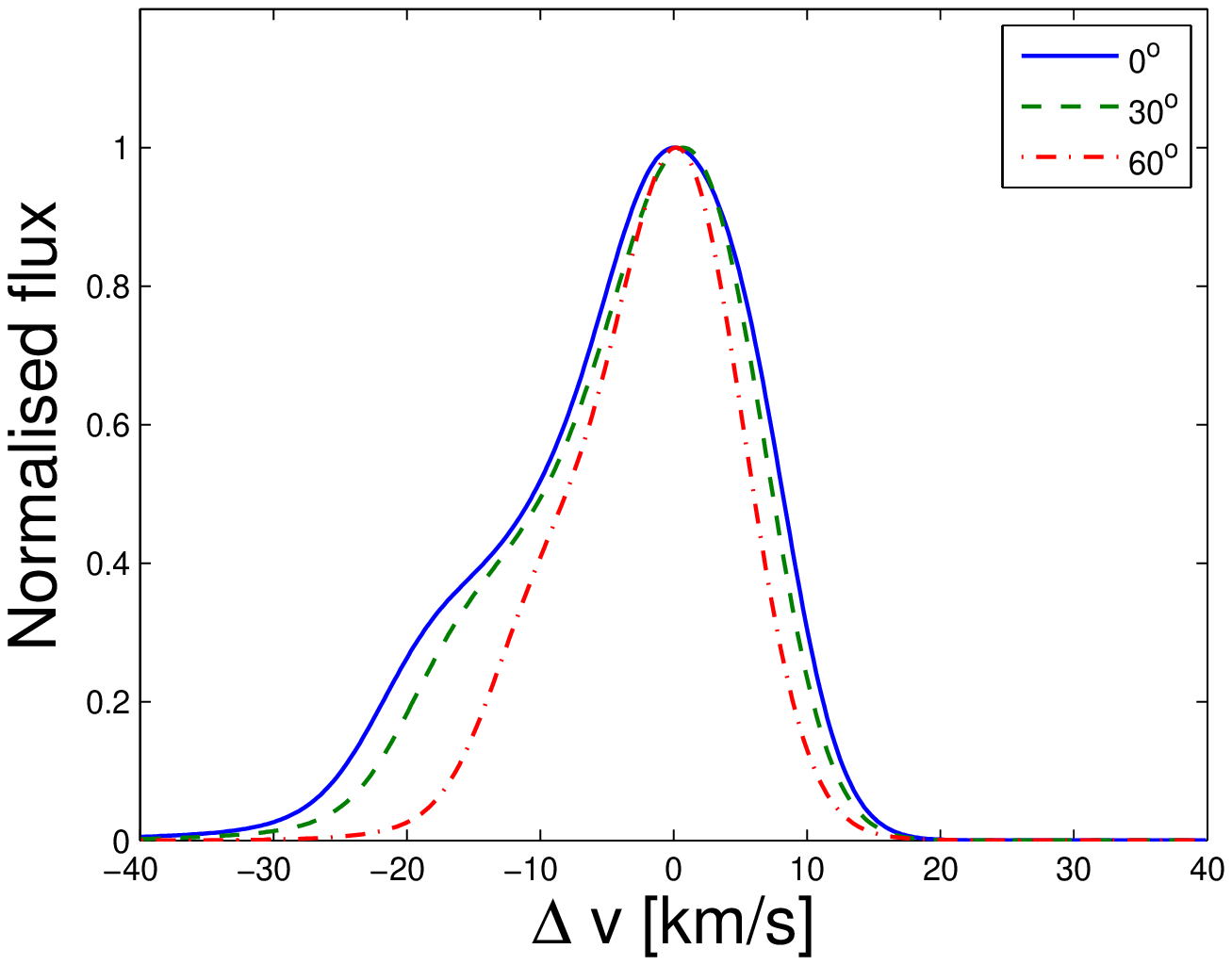}
\caption{There are the profiles of the [Ne II] $12.81 \mu m$ line (left panels), the $H30\alpha$ line (middle panels) and the [Ne III] $15.55 \mu m$ line (right panels) from the cometary H II regions for model F (top panels) and model G (bottom panels).}
\begin{flushleft}
\end{flushleft}
\label{fig_modhi}
\end{figure}

\begin{table}[!htp]\footnotesize
\centering
\begin{tabular}{l|llll|l}
\hline
model F $M_\ast=21.9M_\odot~H=0.05pc$ \\
\hline
Line & Inclination & Peak & FWHM & FWCV & Luminosity \\
 & & ($km~s^{-1}$) & ($km~s^{-1}$) & ($km~s^{-1}$) & ($erg~s^{-1}$) \\
\hline
[Ne II] $12.81 \mu m$ & $0^o$ & -9.3 & 17.0 & -10.48 & $6.07\times10^{34}$ \\
 & $30^o$ & -7.7 & 15.2 & -9.10 \\
 & $60^o$ & -2.5 & 11.6 & -5.27 \\
\hline
$H30\alpha$ & $0^o$ & -10.2 & 37.2 & -11.01 & $5.59\times10^{28}$ \\
 & $30^o$ & -8.6 & 36.4 & -9.54 \\
 & $60^o$ & -5.0 & 35.2 & -5.51 \\
\hline
[Ne III] $15.55 \mu m$ & $0^o$ & -8.5 & 12.6 & -11.33 & $3.36\times10^{34}$ \\
 & $30^o$ & -7.1 & 12.4 & -9.86 \\
 & $60^o$ & -3.1 & 11.2 & -5.73 \\
\hline
model G $M_\ast=21.9M_\odot~v_\ast=10km~s^{-1}$ \\
\hline
Line & Inclination & Peak & FWHM & FWCV & Luminosity \\
 & & ($km~s^{-1}$) & ($km~s^{-1}$) & ($km~s^{-1}$) & ($erg~s^{-1}$) \\
\hline
[Ne II] $12.81 \mu m$ & $0^o$ & 4.3 & 21.0 & -3.82 & $2.92\times10^{35}$ \\
 & $30^o$ & 2.3 & 18.4 & -3.34 \\
 & $60^o$ & 0.5 & 12.8 & -1.93 \\
\hline
$H30\alpha$ & $0^o$ & -1.4 & 38.8 & -3.59 & $2.25\times10^{29}$ \\
 & $30^o$ & -1.4 & 37.6 & -3.11 \\
 & $60^o$ & -1.0 & 35.2 & -1.80 \\
\hline
[Ne III] $15.55 \mu m$ & $0^o$ & 0.1 & 18.4 & -3.82 & $9.46\times10^{34}$ \\
 & $30^o$ & 0.7 & 16.8 & -3.34 \\
 & $60^o$ & 0.3 & 13.6 & -1.94 \\
\hline
\end{tabular}
\caption{Details of the line profiles in model F and model G.\label{tab_modhi}}
\end{table}

\subsection{Line Profiles Computed by Using a Slit}

\begin{figure}[!htp]
\centering
\includegraphics[scale=.24]{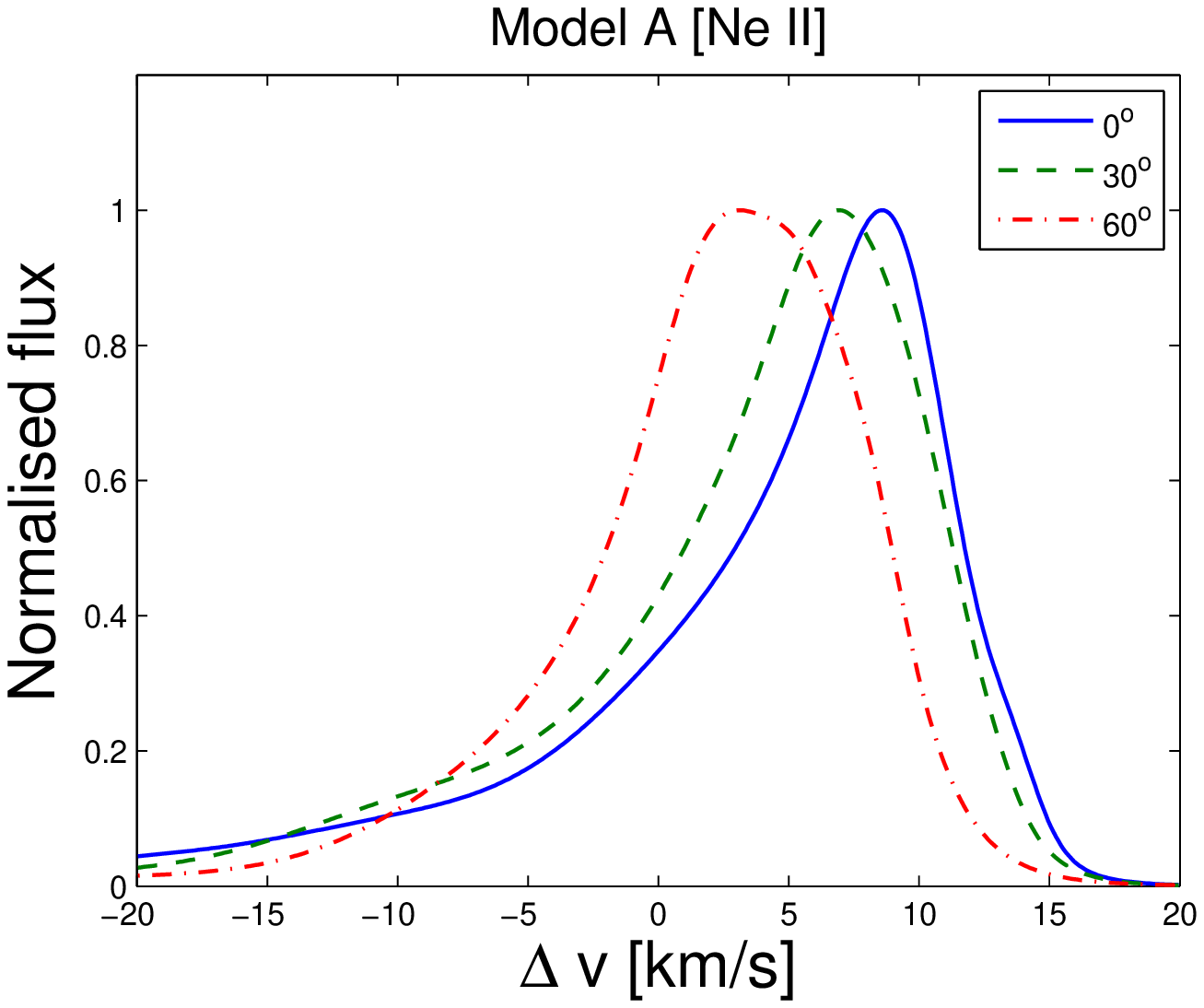}
\includegraphics[scale=.24]{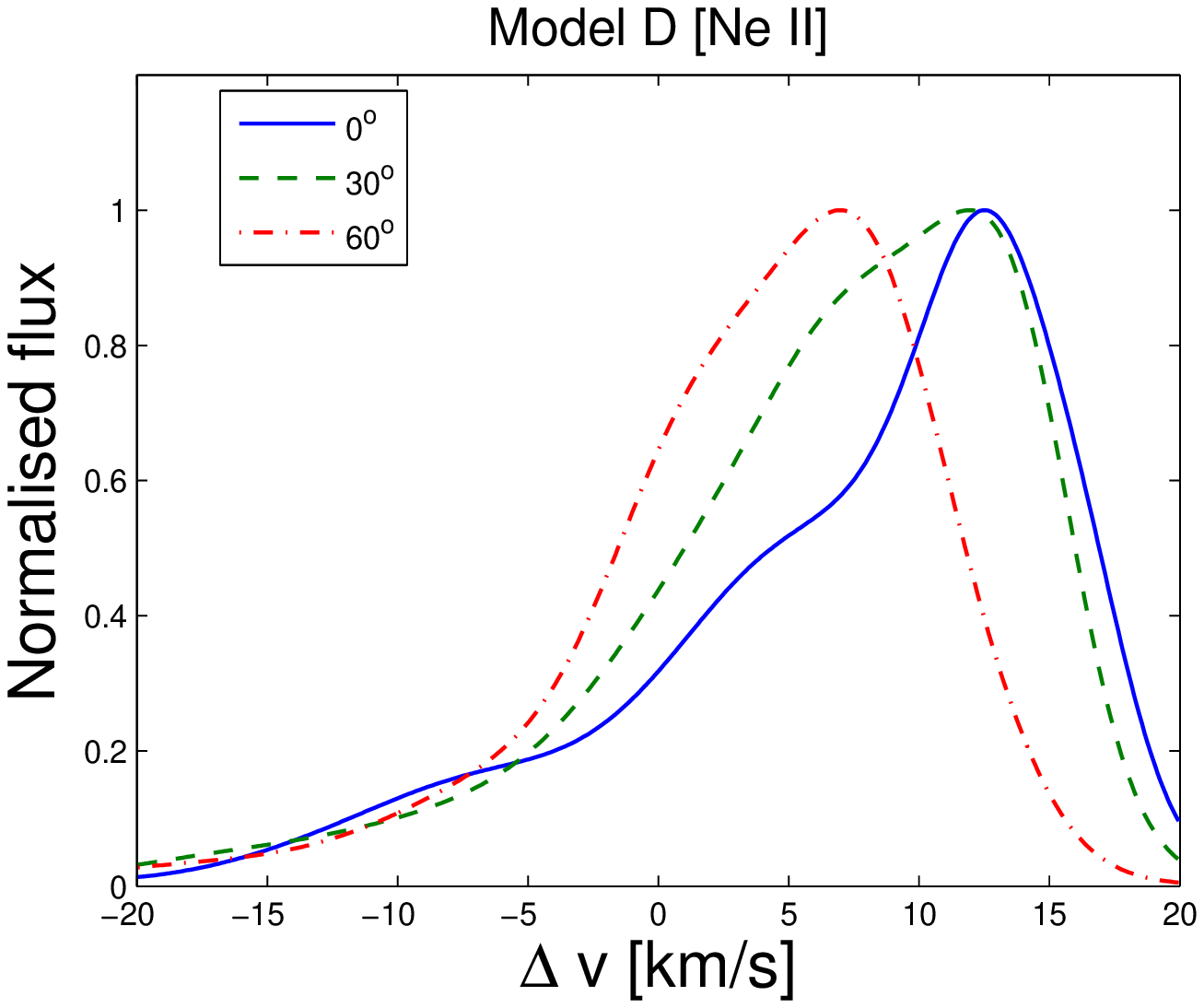}
\includegraphics[scale=.24]{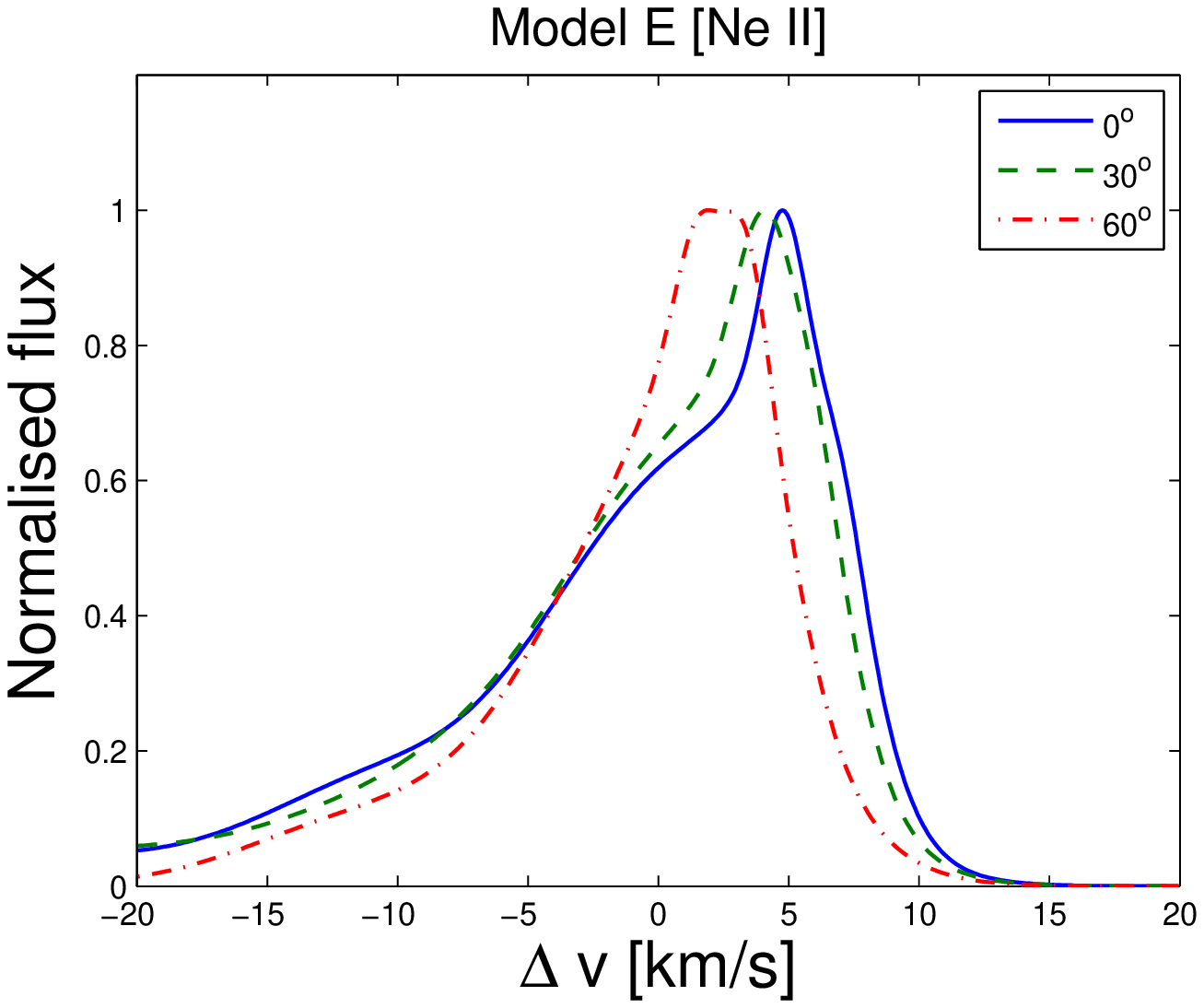}
\includegraphics[scale=.24]{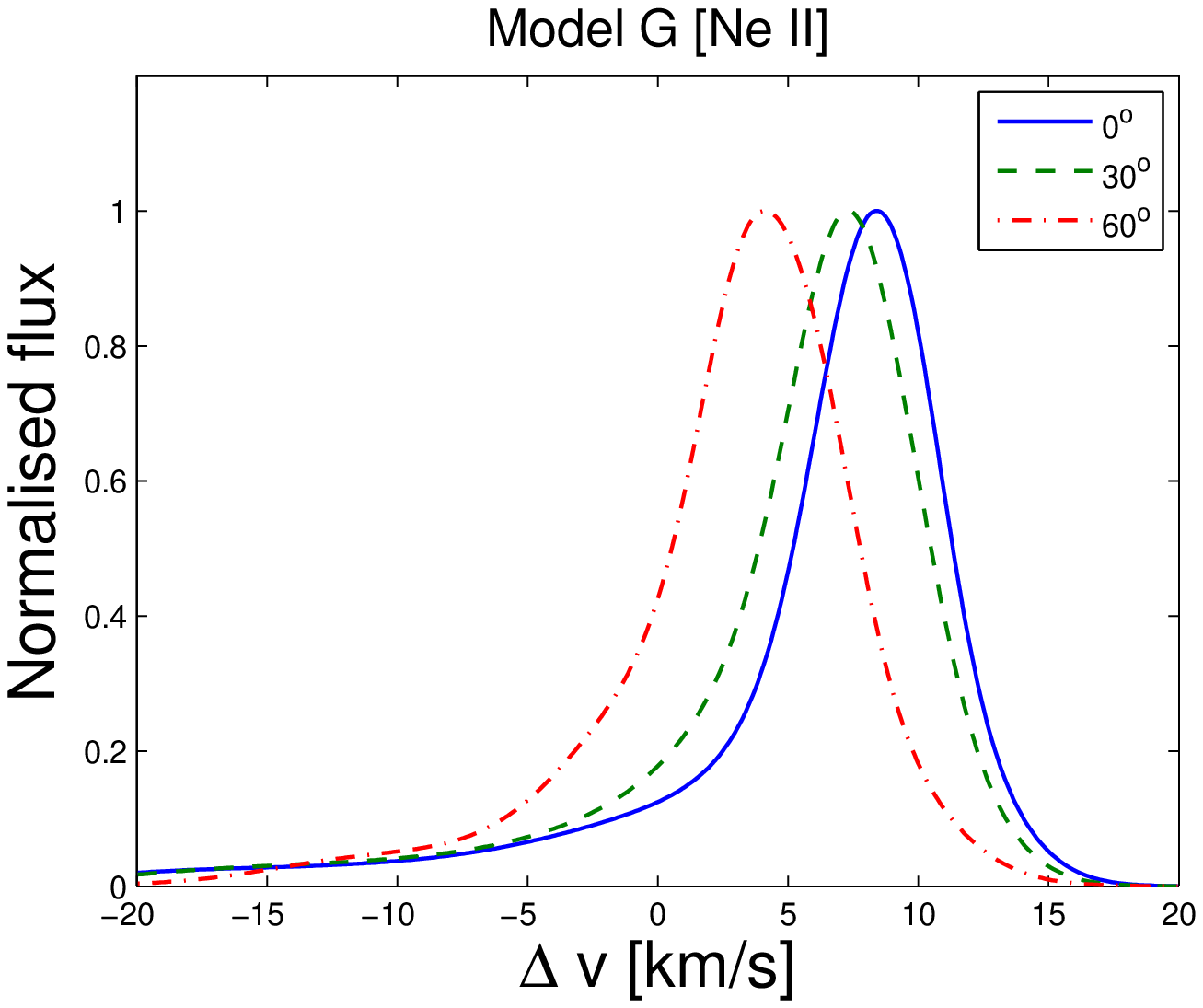}
\includegraphics[scale=.24]{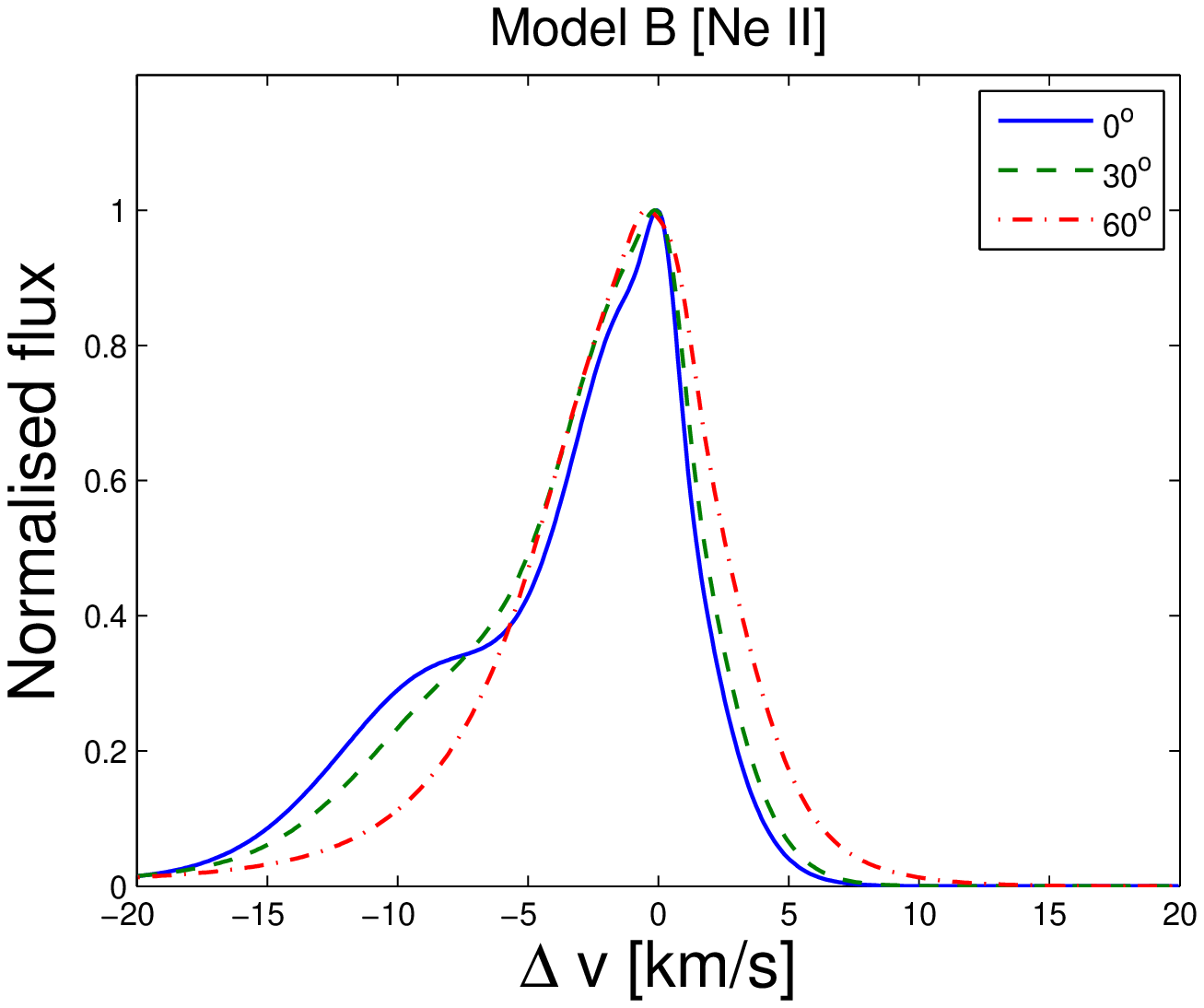}
\includegraphics[scale=.24]{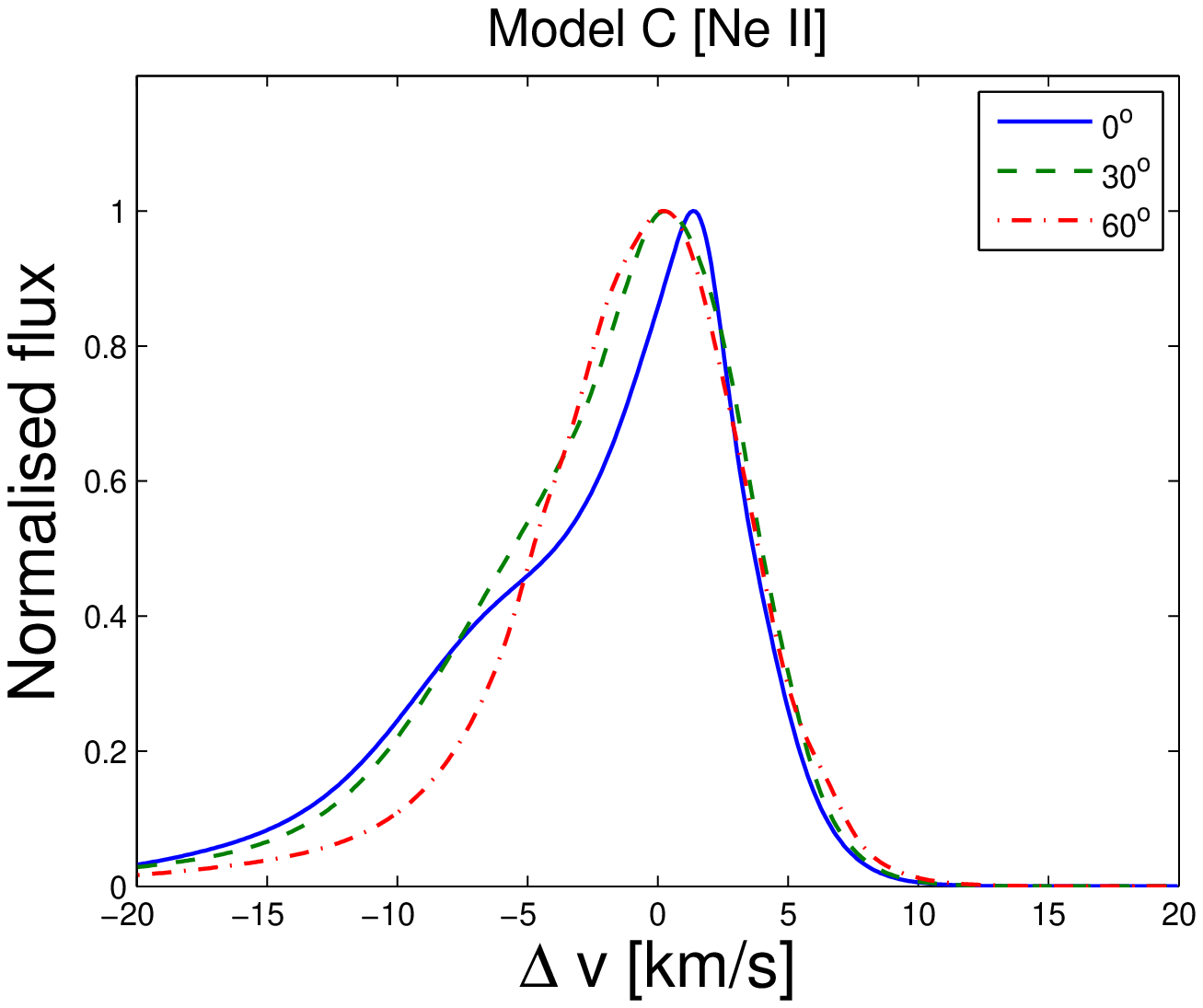}
\includegraphics[scale=.24]{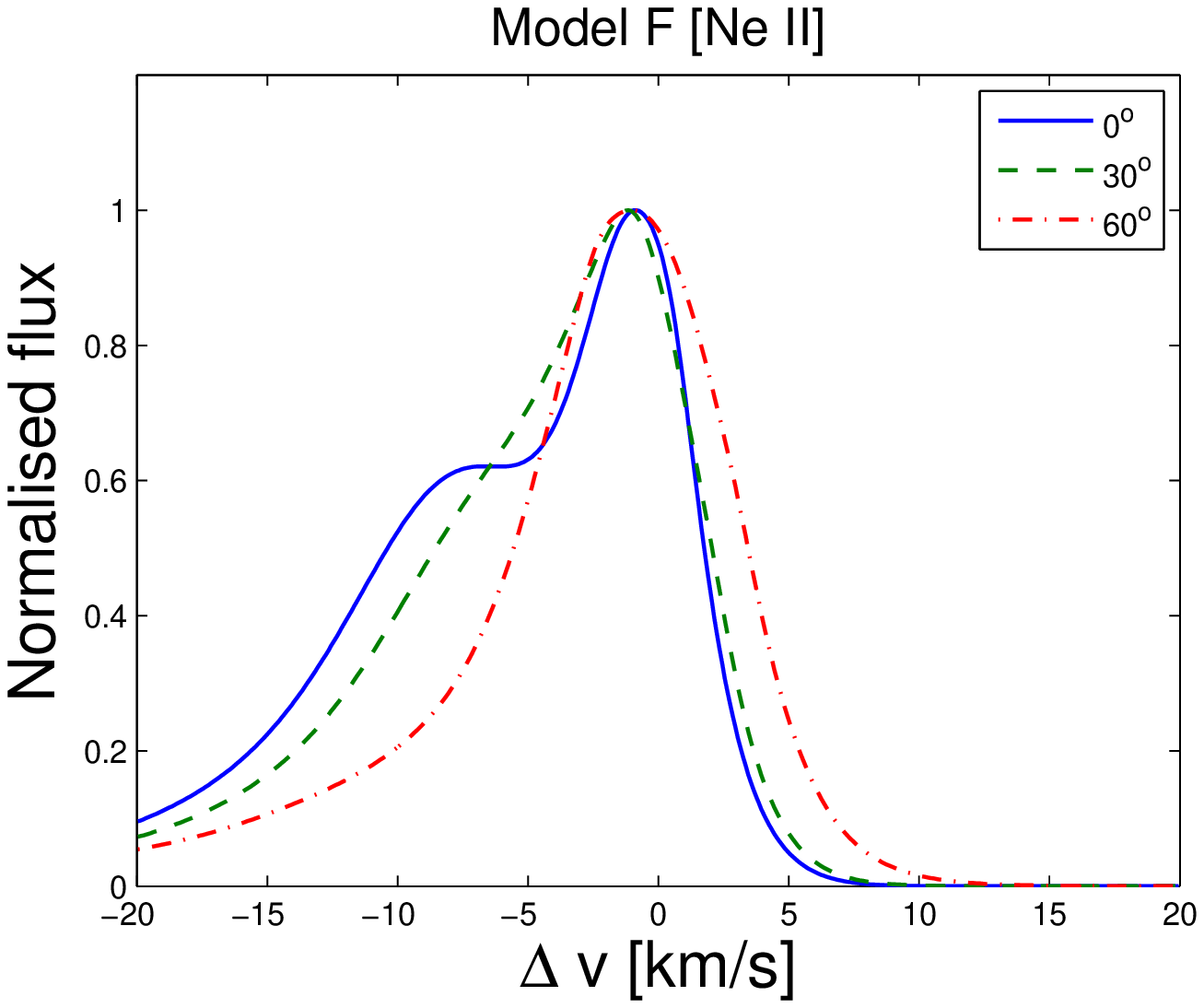}

\caption{There are the profiles of the [Ne II] $12.81\mu m$ line from the slit along the symmetrical axis of the projected 2D image of the models for three inclinations. The top panels are plotted for the bow shock models, and the bottom panels are for the champagne flow models.}
\begin{flushleft}
\end{flushleft}
\label{fig_modslab}
\end{figure}

We have calculated four bow shock models, but only the FWCVs and the peak locations in model D with a high stellar velocity are all red-shifted. Most of the FWCVs and part of the peak locations are blue-shifted because the velocities in the low-density and large-volume inner part of the H II region are always blue-shifted. As is mentioned in \S \ref{sect:stellar motion}, the ionized gas just ionized from the apex of the ached shell has a similar velocity to the stellar velocity and is dense. So if we compute the line profiles from a slit along the symmetrical axis of the projected 2D image rather than from the whole H II region, the influence of the dense gases in the head of the H II region will be highlighted while the emission from the low-density region will be weakened. The method is applied for all the models, and the profiles of the [Ne II] line are presented in Figure \ref{fig_modslab}. The peak locations of the profiles of the [Ne II] line and the $H30\alpha$ line are shown in Table \ref{tab_modslab}.

For the [Ne II] line, the profiles are all single-peaked. In bow shock models (model A, D, E and G), the peak locations are slightly less than $cos(\theta)v_*$. In champagne flow models (model B, C and F), it is different that the peak locations of the [Ne II] line profiles are close to zero. The red-shifted peak locations in model C suggest that the propagation velocity of the ionization front in model C is a little faster than in other champagne flow models at the age of $160,000~yr$. For the $H30\alpha$ line, because of the large thermal broadening, the relation of the peak locations to the stellar velocities in bow shock models is less explicit. But, the champagne flow models are easily distinguished from the bow shock models due to the red-shifted peak locations in bow shock models and the blue-shifted values in champagne flow models.

When comparing the FWCVs presented in Table \ref{tab_modslab}, we find that the FWCVs in champagne flow models are all blue-shifted. In bow shock models, the FWCVs are obviously red-shifted when the stellar velocity is high. But, the FWCVs in model E are slightly blue-shifted due to the low stellar velocity ($v_*=5~km~s^{-1}$). And it suggests that the FWCVs in a bow shock model with a stellar velocity lower than $5~km~s^{-1}$ are not easily distinguished from the corresponding values in champagne flow models.

\begin{table}[!htp]\footnotesize
\centering
\begin{tabular}{l|lll|lll}
\hline
 Peak($km~s^{-1}$) & [Ne II] & $12.81\mu m$ &  & $H30\alpha$ & & \\
\hline
Inclination & $0^o$ & $30^o$ & $60^o$ & $0^o$ & $30^o$ & $60^o$   \\
\hline
Model A & 8.6 & 7.0 & 3.1 & 7.8 & 6.6 & 3.8 \\
Model B & -0.1 & -0.2 & -0.5 & -2.6 & -2.2 & -1.4 \\
Model C & 1.4 & 0.3 & 0.3 & -1.4 & -1.4 & -0.6 \\
Model D & 12.6 & 12.0 & 7.0 & 8.6 & 7.4 & 4.2 \\
Model E & 4.8 & 4.1 & 2.0 & 2.6 & 2.2 & 1.0 \\
Model F & -0.9 & -1.2 & -1.1 & -3.0 & -2.6 & -1.4 \\
Model G & 8.5 & 7.4 & 4.2 & 7.4 & 6.2 & 3.4 \\
\hline
 FWCV($km~s^{-1}$) & [Ne II] & $12.81\mu m$ &  & $H30\alpha$ & & \\
\hline
Inclination & $0^o$ & $30^o$ & $60^o$ & $0^o$ & $30^o$ & $60^o$   \\
\hline
Model A & 4.18 & 3.50 & 2.07 & 4.92 & 4.26 & 2.46 \\
Model B & -3.82 & -3.24 & -1.77 & -4.95 & -4.29 & -2.47 \\
Model C & -2.42 & -2.09 & -1.24 & -4.29 & -3.72 & -2.15 \\
Model D & 7.27 & 6.46 & 3.86 & 6.14 & 5.32 & 3.07 \\
Model E & -0.12 & -0.30 & -0.57 & -1.00 & -0.86 & -0.50 \\
Model F & -5.38 & -4.60 & -2.60 & -5.70 & -4.94 & -2.85 \\
Model G & 6.08 & 5.18 & 2.89 & 5.16 & 4.47 & 2.58 \\
\hline
\end{tabular}
\caption{The peak locations and the flux weighted central velocities of the [Ne II] $12.81\mu m$ line and the $H30\alpha$ line from the slit along the symmetrical axis of the projected 2D image.\label{tab_modslab}}
\end{table}

\subsection{the Flux Weighted Central Velocities}
\label{sect:FWCV}

As is mentioned in \S \ref{sect:model B}, the number density is $n \approx n_e \ll n_{cr}$. And in our calculation, the temperature is always close to the equilibrium temperature in photoionized region. So if we assumed $T\approx10000K$ in photoionized region, the flux of the [Ne II] and the [Ne III] line from a unit volume is approximately as follow:

\begin{equation}
f_{ul}(i)=\begin{cases}
b_kN_kA_{ul}h\nu_{ul} & for~H30\alpha,~k=30,~i=0  \\
Ab_{Ne}X(Ne^+)(n_e^2/(2n_{cr}))A_{ul}h\nu_{ul} & for~\textrm{[Ne II]},~i=1 \\
Ab_{Ne}X(Ne^{2+})(5n_e^2/(3n_{cr1}))A_{ul}h\nu_{ul} & for~\textrm{[Ne III]},~i=2

\end{cases}~~~~.
\label{equ_ful}
\end{equation}

Since the neon atoms are almost all ionized in the H II region, we can get the relation that $X(Ne^+)+X(Ne^{2+})=1$. Other parameters in Eq. (\ref{equ_ful}) are constant due to the constant temperature in photoionized region. Hence, the fluxes from a unit volume of the three lines are all proportional to $n_e^2$. Then we can obtain the following relation:

\begin{equation}
(\frac{L(1)}{1.302\times10^{-21}}\bar{v}(1)+\frac{L(2)}{1.162\times10^{-21}}\bar{v}(2))/(\frac{L(1)}{1.302\times10^{-21}}+\frac{L(2)}{1.162\times10^{-21}})=\bar{v}(0)~~,
\end{equation}

where $L(1)$ and $L(2)$ are the line luminosities of the [Ne II] line and the [Ne III] line, respectively. $\bar{v}(0)$, $\bar{v}(1)$ and $\bar{v}(2)$ are the FWCVs of the profiles of the $H30\alpha$ line, the [Ne II] line and the [Ne III] line, respectively. If the abundance of neon has been obtained, we can approximately compute the line luminosity and the FWCVs of the [Ne III] line from those of the other two lines:

\begin{equation}
L(2)=(L(0)-\frac{L(1)}{1.801\times10^{10}Ab_{Ne}})\times1.608\times10^{10}Ab_{Ne}~~~~,
\label{equ_lim}
\end{equation}

\begin{equation}
\bar{v}(2)=[\bar{v}(0)(\frac{L(1)}{1.302\times10^{-21}}\bar{v}(1)+\frac{L(2)}{1.162\times10^{-21}}\bar{v}(2))-\frac{L(1)}{1.302\times10^{-21}}\bar{v}(1)]\frac{1.162\times10^{-21}}{L(2)}~~~~,
\label{equ_fwcv}
\end{equation}

where $L(0)$ is the line luminosity of the $H30\alpha$ line. The [Ne III] line is not detectable from the ground, but the [Ne II] line and the $H30\alpha$ line can be observed with the ground based telescopes. These equations could be useful since the difference in the [Ne III] line between the bow shock models and the champagne flow models is the most obvious among the three lines. For observing the [Ne III] line, a typical resolution ($\lambda/\Delta\lambda=30,000$) is needed to resolve the difference in the [Ne III] line profiles. The line also needs to be observed by using space telescopes since it can not be observed from the ground. In our models, the average deviation of the approximate values computed with the Eq. (\ref{equ_lim}) and (\ref{equ_fwcv}) from the accurate values is 6\%.

\section{Conclusions}

When considering the gas kinematics of compact H II region models, people generally predict that the line emission from the ionized region may be red-shifted along the line of sight for a bow shock and blue-shifted for a champagne flow when looking into the regions from the tail. Our simulation proves that this envision is generally true, but it is violated in the case of low/medium stellar speed bow shock. When the stellar velocity is lower than $5~km~s^{-1}$, it is difficult to distinguish the bow shock from a champagne flow. On the contrary, A bow shock is very easily distinguished from champagne flows if the stellar velocity is higher than $10~km~s^{-1}$.

In this paper, we have simulated the evolution of the cometary H II region by bow shock models and champagne flow models, and have displayed the density distribution and velocity fields in bow shock models and champagne flow models. We have studied the [Ne II] $12.81\mu m$ line, the $H30\alpha$ line and the [Ne III] $15.55\mu m$ line profiles from these models. We make comparisons of the line profiles in bow shock models and champagne flow models. We find that:

1. After the champagne models reach quasi-steady state, the line profiles are all biased towards the blue-shifted side and have obviously blue-shifted peak locations and flux weighted central velocities. The peak locations and the flux weighted central velocities of the [Ne II] line profiles are the least blue-shifted in the three lines meanwhile those of the [Ne III] line profiles are the most blue-shifted in champagne flow models. In our champagne flow models, when the inclination angle is no more than $60^o$, the flux weighted central velocities are lower than $-3.0~km~s^{-1}$ for the [Ne II] line, are lower than $-4.0~km~s^{-1}$ for the $H30\alpha$ line and are lower than $-4.5~km~s^{-1}$ for the [Ne III] line.

2. For bow shock models, We find that the line profiles become more biased towards the red-shifted side with increasing stellar velocity. Except for the model D with a star of $v_\ast=5~km~s^{-1}$, in the bow shock models computed in this paper, the flux weighted central velocities are higher than $-3.82~km~s^{-1}$ for the [Ne II] line and the [Ne III] line, and are higher than $-3.6~km~s^{-1}$ for the $H30\alpha$ line.


3. The peak locations of the [Ne II] line profiles from the slits along the symmetrical axis of the projected 2D image are slightly less than $cos(\theta)v_\ast$ in bow shock models and no more than $1.5~km~s^{-1}$ in champagne flow models. The peak locations of the $H30\alpha$ line profiles from the slits are less red-shifted or more blue-shifted than those of the [Ne II] line.

We have shown the profiles of the [Ne II] line at $12.81 \mu m$, the $H30\alpha$ recombination line and the [Ne III] line at $15.55 \mu m$ in this paper. These results are useful to recognizing the causes of the cometary H II region.

\normalem
\begin{acknowledgements}
The work is partially funded by National Basic Research Program of China (973 program) No. 2012CB821805. The authors are also grateful to the supports by Doctoral Fund of Ministry of Education of China No. 20113402120018 and Natural Science Foundation of Anhui Province of China No. 1408085MA13. F.-Y. Zhu is thankful to Dr. X.-L. Deng for generous financial aids during the visit to Beijing Computational Science Research Center and suggestions on the hydrodynamical computation methods. F.-Y. Zhu would also thank Dr. J.-X. Wang for financial supports.

\end{acknowledgements}

\bibliographystyle{raa}
\bibliography{bibtex}

\begin{thebibliography}{37}
\providecommand{\natexlab}[1]{#1}
\providecommand{\selectlanguage}[1]{\relax}

\bibitem[Alexander(2008)]{ale08} Alexander, R. D. 2008, \mnras, 391, L64

\bibitem[Arthur \& Hoare(2006)]{art06} Arthur, S. J., \& Hoare, M. G. 2006, \apjs, 165, 283

\bibitem[Bakes \& Tielens(1994)]{bak94} Bakes, E. L. O., \& Tielens, A. G. G. M. 1994, \apj, 427, 822

\bibitem[Bodenheimer et al.(1979)]{bod79} Bodenheimer, P., Tenorio-Tagle, G., \& Yorke, H. W. 1979, \apj, 233, 85

\bibitem[Comer\'{o}n(1997)]{com97} Comer\'{o}n, F. 1997, \aap, 326, 1195

\bibitem[Comer\'{o}n \& Kaper(1998)]{com98} Comer\'{o}n, F., \& Kaper, L. 1998, \aap, 338, 273

\bibitem[Dale et al.(2013)]{dal13} Dale, J. E., Ngoumou, J., Ercolano, B., \& Bonnell, I. A. 2013, \mnras, 436, 3430

\bibitem[Diaz-Miller et al.(1998)]{dia98} Diaz-Miller, R. I., Franco, J., \& Shore, S. N. 1998, \apj, 501, 192

\bibitem[Draine \& Bertoldi(1996)]{dra96} Draine, B. T., \& Bertoldi, F. 1996, \apj, 468, 269

\bibitem[Gendelev \& Krumholz(2012)]{gen12} Gendelev, L., \& Krumholz, M. R. 2012, \apj, 745, 158

\bibitem[Glassgold et al.(2007)]{gla07} Glassgold, A. E., Najita, J. R., Igea, J. 2007, \apj, 656, 515

\bibitem[Henry(1970)]{hen70} Henry, R. J. W. 1970, \apj, 161, 1153

\bibitem[Hibbert \& Scott(1994)]{hib94} Hibbert, A., \& Scott, M. P. 1994, J. phys. B: At. Mol. Opt. phys., 27, 1315

\bibitem[Hollenbach \& McKee(1979)]{hol79} Hollenbach, D., \& McKee, C. F. 1979, \apj, 41, 555

\bibitem[Hollenbach \& McKee(1989)]{hol89} Hollenbach, D. J., \& McKee, C. F. 1989, \apj, 342, 306

\bibitem[Hollenbach \& Tielens(1999)]{hol99} Hollenbach, D. J., \& Tielens, A. G. G. M. 1999, Rev. Mod. Phys., 71, 173

\bibitem[Holweger(2001)]{hol01} Holweger, H. 2001, AIP Conf. Proc. 598, 23

\bibitem[Hosokawa \& Inutsuka(2006)]{hos06} Hosokawa, T., \& Inutsuka, S. 2006, \apj, 646, 240

\bibitem[Israel(1978)]{isr78} Israel, F. P. 1978, \aap, 70, 769

\bibitem[Kurtz et al.(1994)]{kur94} Kurtz, S., Churchwell, E., \& Wood, D. O. S. 1994, \apjs, 91, 659

\bibitem[Lee et al.(1996)]{lee96} Lee, H.-H., Herbst, E., Pineau des For\^{e}ts, G., Roueff, E., Le Bourlot, J. 1996, \aap, 311, 690L

\bibitem[Mac Low et al.(1991)]{mac91} Mac Low, M.-M., van Buren, D., Wood, D. O. S., \& Churchwell, E. 1991, \apj, 369, 395

\bibitem[Mellema \& Lundqvist(2002)]{mel02} Mellema, G., \& Lundqvist, P. 2002, \aap, 394, 901

\bibitem[Miyoshi \& Kusano(2005)]{miy05} Miyoshi, T., \& Kusano, K. 2005, J. Comput. Phys., 208, 315

\bibitem[Morisset et al.(2002)]{mor02} Morisset, C., Schaerer, D., et al. 2002, \aap, 386, 558M

\bibitem[Nelson \& Langer(1997)]{nel97} Nelson, R. P., \& Langer, W. D. 1997, \apj, 482, 796

\bibitem[P\'{e}quignot et al.(1991)]{peq91} P\'{e}quignot, D., Petitjean, P. \& Boisson, C. 1991, \aap, 251, 680

\bibitem[Raga et al.(1997)]{rag97} Raga, A. C., Noriega-Crespo, A., et al. 1997, Rev. Mex. AA, 33, 73

\bibitem[Seaton(1959)]{sea59} Seaton, M. F. 1959, \mnras, 119, 90S

\bibitem[Spitzer(1978)]{spi78} Spitzer, L. 1978, Physical Processes in the Interstellar Medium (New York: Wiley)

\bibitem[Str\"{o}mgren(1939)]{str39} Str\"{o}mgren, B. 1939, \apj, 89, 526

\bibitem[Tenorio-Tagle(1979)]{ten79} Tenorio-Tagle, G. 1979, \aap, 71, 59

\bibitem[Tielens(1985)]{tie85} Tielens, A. G. G. M., \& Hollenbach, D. 1985, \apj, 291, 722

\bibitem[van Buren \& Mac Low(1992)]{bur92} van Buren, D., \& Mac Low, M.-M. 1992, \apj, 394, 534

\bibitem[Walsh et al.(1998)]{wal98} Walsh, A. J., Burton, M. G., Hyland, A. R., \& Robinson, G. 1998, \mnras, 301, 640

\bibitem[Wilkin(1996)]{wil96} Wilkin, F.P. 1996, \apjl, 459, L31

\bibitem[Wood \& Churchwell(1989)]{woo89} Wood, D. O. S., \& Churchwell, E. 1989, \apjs, 69, 831

\bibitem[Yorke et al.(1983)]{yor83} Yorke, H. W., Tenorio-Tagle, G., \& Bodenheimer, P. 1983, \aap, 127, 313

\end{thebibliography}

\end{document}